\documentclass[aps,prb,twocolumn,superscriptaddress]{revtex4-2}

\usepackage{graphicx}
\usepackage{dcolumn}
\usepackage{bm}
\usepackage{color}
\usepackage{amsmath}
\usepackage[normalem]{ulem}

\begin{document}

\preprint{APS/123-QED}

\title{Chemical potential of magnetic skyrmion quasiparticles in heavy metal/iron 
bilayers}

\author{Bal\'azs Nagyfalusi}
 \email{nagyfalusi.balazs@ttk.bme.hu}
\affiliation{
Department of Theoretical Physics, Institute of Physics, Budapest University of Technology and Economics, M\H uegyetem rkp. 3, H-1111 Budapest, Hungary}

\author{ L\'aszl\'o Udvardi}%
\affiliation{
Department of Theoretical Physics, Institute of Physics, Budapest University of Technology and Economics, M\H uegyetem rkp. 3, H-1111 Budapest, Hungary}

\author{ L\'aszl\'o Szunyogh}%
\affiliation{
Department of Theoretical Physics, Institute of Physics, Budapest University of Technology and Economics, M\H uegyetem rkp. 3, H-1111 Budapest, Hungary}
\affiliation{
HUN-REN-BME Condensed Matter Research Group, Budapest University of Technology and Economics, M\H uegyetem rkp. 3, H-1111 Budapest, Hungary
}

\author{Levente R\'ozsa}%
\affiliation{
Department of Theoretical Solid State Physics, Institute for Solid State Physics and Optics, HUN-REN Wigner Research Center for Physics,  H-1525 Budapest, Hungary
}
\affiliation{
Department of Theoretical Physics, Institute of Physics, Budapest University of Technology and Economics, M\H uegyetem rkp. 3, H-1111 Budapest, Hungary}

\date{\today}

\begin{abstract}
We performed metadynamics Monte Carlo simulations to obtain the free energy as a function of the topological charge in the skyrmion-hosting magnetic model systems (Pt$_{0.95}$Ir$_{0.05}$)/Fe/Pd(111) and Pd/Fe/Ir(111), using a spin model containing parameters based on {\em ab initio} calculations. 
Using the topological charge as collective variable, this method allows for evaluating the temperature dependence of the number of skyrmionic quasiparticles. 
In addition, from the free-energy cost of increasing and decreasing the topological charge of the system we determined
chemical potentials as a function of the temperature. 
At lower temperature, the chemical potential for creating skyrmions and antiskyrmions from the topologically trivial state is different. This splitting of the chemical potential is particularly pronounced for large external magnetic fields when the system is in a field-polarized phase. 
We observed a change in the shape of the free-energy curves when skyrmion-skyrmion interactions become more pronounced.
\end{abstract}

\maketitle

\section{\label{sec:introduction} Introduction}

Since their first experimental indications for the formation of a skyrmion lattice in bulk MnSi~\cite{Muhlbauer2009,Neubauer2009}
and  its  real-space  imaging  on Fe$_{0.5}$Co$_{0.5}$Si film using Lorentz transmission electron microscopy~\cite{Yu2010},
magnetic skyrmions have been  observed in a wide variety of  
bulk materials and 
thin films~\cite{Tokura2021}.  Skyrmions  
have received 
 very broad  attention since they hold great promise in the development of spintronic devices~\cite{Back_2020}. These whirling magnetic patterns can be stabilized by the Dzyaloshinky-Moriya interaction (DMI) \cite{DZYALOSHINSKY1958,Moriya1960, Bogdanov1994}, by four-spin interactions \cite{Heinze2011}, or due to the frustration of Heisenberg exchange interactions \cite{Okubo2012,vonMalottki2017}.
Their special properties, like the exceptional
stability,  the nanoscale size 
and the considerable mobility, make magnetic skyrmions perfect candidates for usage in logical and high-density memory devices \cite{Fert2013,Iwasaki2013,Zhou2014}, and they form an emerging platform for future neuromorphic~\cite{Grollier2020} and quantum computing~\cite{Psaroudaki2021}.

In the present study we focus on two kinds of  skyrmion-hosting  ultrathin film systems, namely a Pd/Fe bilayer on Ir(111) \cite{Romming2013,Romming2015, Rozsa2016,Bottcher2018}, and a Pt$_{0.95}$Ir$_{0.05}$/Fe bilayer on Pd(111) \cite{Rozsa2016PRL,Rozsa2017,Schick2021}.
In the case of Pd/Fe/Ir(111), Dzyaloshinky-Moriya interactions competing with isotropic Heisenberg exchange, uniaxial anisotropy and Zeeman terms lead to the formation of 
 non-collinear  spin structures. Spin-polarized scanning tunneling microscopy experiments~\cite{Romming2013} revealed that the ground state of the system is a spin-spiral state.  Applying an  external magnetic field perpendicular to the  surface  results in the coexistence of spin spirals and N\'eel-type skyrmions. Increasing the external field above 1~T, the spin spirals disappear, and a 
hexagonal skyrmion lattice emerges. A further increase of the magnetic field brings the system into a field-polarized (FP) phase. 
The experimental results have been reproduced by Monte Carlo simulations based on parameters obtained from first-principles density-functional theory calculations~\cite{Rozsa2016,Bottcher2018,Miranda2022}. Magnetic skyrmions are 
 characterized  by their topological charge $Q$,  which counts how many times the spin configuration wraps around the unit sphere upon the application of a stereographic projection. 
 The interplay between frustrated Heisenberg
exchange and DMI can lead to the formation of metastable skyrmionic spin structures with various topological charges, which have been theoretically investigated in (Pt$_{0.95}$Ir$_{0.05}$)/Fe/Pd(111)~\cite{Rozsa2017,Weissenhofer2019,Rozsa2020,Weissenhofer2021}. 

The topological charge cannot be changed dynamically in a continuum model, being the reason why skyrmions are called topologically protected. 
However,
 on the discretized atomic level  the topological charge is not perfectly conserved. Overcoming the finite energy barrier between the skyrmionic and topologically trivial states allows to create and annihilate skyrmions, which can then be regarded as quasiparticles~\cite{Schick2021}.  The energy barriers and the attempt frequencies of skyrmion creation and annihilation have been thoroughly investigated based on the zero-temperature energy landscape~\cite{BESSARAB2015335,vonMalottki2017,Cortes-Ortuno2017,Desplat2018,vonMalottki2019}, with the results comparing favourably with the collapse mechanisms observed with low-temperature scanning tunneling microscopy~\cite{Muckel2021}. Skyrmion lifetimes have also been calculated based on Metropolis Monte Carlo~\cite{Hagemeister2015} and atomistic spin dynamics~\cite{Rozsa2016} simulations, which are primarily applicable at higher temperatures where the creation and annihilation processes are faster. Deviations from the method based on the zero-temperature energy landscape have been observed in this elevated temperature range~\cite{Schick2021}. Bridging the low-temperature and high-temperature limits can be achieved by speeding up the numerical simulations using specialized algorithms. A path-sampling approach was applied for this purpose in Ref.~\cite{Desplat2020}, and the finite-temperature free-energy barriers of single skyrmions were also recently investigated in Ref.~\cite{charalampidis2023arxiv} 
using the metadynamics method~\cite{Bussi2020}.
 Most of these previous studies focused on the stability of single isolated skyrmions. Experimental observations in Ref.~\cite{Lindner2020} indicated that at elevated temperatures multiple skyrmions may be created or annihilated at at varying positions in nanoislands. It is expected that skyrmion-skyrmion interactions in such ensembles influence the stability of the quasiparticles, as the influence of interactions has already been demonstrated on the thermally induced motion of skyrmions~\cite{Schaeffer2019,Song2021,Ge2023}.

Here,   we explore the free-energy landscape of larger skyrmion ensembles with various topological charges in the Pd/Fe/Ir(111) and Pt$_{0.95}$Ir$_{0.05}$/Fe/Pd(111) systems also by performing  metadynamics simulations 
implemented with the Metropolis Monte Carlo algorithm. 
In metadynamics the free energy of the system is calculated as a function of a collective variable appropriate to the problem under investigation.
Choosing the topological  charge $Q$ as collective variable we are able to determine the equilibrium  number of skyrmions as a function of the temperature, and to obtain the temperature dependent chemical potential of  skyrmions and antiskyrmions. 

The paper is structured as follows. In Section~\ref{sec:methods} we briefly introduce the spin model based on first-principles parameters,  and the details of the metadynamics Monte Carlo simulations applied to magnetic skyrmions are described in detail. 
In Section~\ref{sec:results} we present our results for the  free-energy landscapes in the  (Pt$_{0.95}$Ir$_{0.05}$)/Fe/Pd(111) and the Pd/Fe/Ir(111) bilayer systems, which are in good agreement with the 
 conclusions concerning skyrmion stability in these systems  published in Refs.~\cite{Rozsa2016} and \cite{Schick2021}  based on atomistic simulations . In Section~\ref{sec:conclusion} we draw conclusions underpinning the quasiparticle picture of magnetic skyrmions.

\section{Methods}
\label{sec:methods}

\subsection{Spin models}

The magnetic structure of the thin films is studied in terms of a classical atomistic Heisenberg model,
\begin{eqnarray} \label{eq:H}
	   H &=& - \frac{1}{2}\sum_{i\neq j} \mathbf{s}_{i}^T\mathbf{J}_{ij}\mathbf{s}_{j}  
      +\sum_i\lambda_{z} (\mathbf{s}_{i}\hat{\mathbf{z}} )^2
     - \mu \sum_i  \mathbf{s}_i\mathbf{B} 
      \;, 
 \end{eqnarray}
where the sums run over sites of the Fe atomic layer, $\mathbf{s}_{i}$ is a unit vector representing the direction of the atomic spin magnetic moment at site $i$ and $\mathbf{J}_{ij}$ is a $3\times 3$ tensor of exchange interactions which can be decomposed as~\cite{Udvardi2003} 
\begin{eqnarray}
\mathbf{J}_{ij}&=&\frac{1}{3}\textrm{Tr}\left(\mathbf{J}_{ij}\right)\mathbf{I}+\frac{1}{2}\left(\mathbf{J}_{ij}-\mathbf{J}^{T}_{ij}\right)\nonumber
\\
& \ &+\frac{1}{2}\left(\mathbf{J}_{ij}+\mathbf{J}^{T}_{ij}-\frac{2}{3}\textrm{Tr}\left(\mathbf{J}_{ij}\right)\mathbf{I}\right),\label{Jij}
\label{eq:Jij}
\end{eqnarray}
where $\frac{1}{3}\textrm{Tr}\left(\mathbf{J}_{ij}\right)=J_{ij}$ is the isotropic Heisenberg exchange coupling, $\mathbf{s}^{T}_{i}\frac{1}{2}\left(\mathbf{J}_{ij}-\mathbf{J}^{T}_{ij}\right)\mathbf{s}_{j}=\mathbf{D}_{ij}\left(\mathbf{s}_{i}\times\mathbf{s}_{j}\right)$ stands for the antisymmetric Dzyaloshinky-Moriya interaction, and the last, symmetric traceless part in Eq. (\ref{eq:Jij}) includes 
the pseudo-dipolar anisotropy contributions induced by the spin-orbit coupling. The on-site uniaxial magnetocrystalline anisotropy is characterized by the constant $\lambda_{z}$, where $\hat{\mathbf{z}} $ is a unit vector  parallel to the $[111]$ direction. Finally $\mu$ is the magnitude of the Fe spin magnetic moments, and $\mathbf{B}=B\,\hat{\mathbf{z}} $ is an applied external magnetic field perpendicular to the surface.

The  parameters of the spin models  for the (Pt$_{0.95}$Ir$_{0.05}$)/Fe/Pd(111) and Pd/Fe/Ir(111) bilayers are taken from Refs. \cite{Rozsa2016PRL} and \cite{Simon2014}, respectively. 
The obtained spin models confirmed a cycloidal spin-spiral ground state for both systems.  This transforms at $B=0.21~\textrm{T}$ directly into the field-polarized state where isolated skyrmions can be observed in (Pt$_{0.95}$Ir$_{0.05}$)/Fe/Pd(111), while the skyrmion lattice is a stable thermodynamical phase between $B=1.4~\textrm{T}$ and $B=3.0~\textrm{T}$ in Pd/Fe/Ir(111). 

\subsection{Metadynamics simulations}

The finite-temperature behaviour of the magnetic systems were studied
by using well-tempered metadynamics simulations~\cite{Laio2002, Barducci2008}. 
This method has recently been successfully implemented to study the magnetic anisotropy energy of thin films~\cite{Nagyfalusi2019, Nagyfalusi2020}. While the details of the computational scheme are given in these references, next we describe the advances of the method in simulating magnetic skyrmions at finite temperatures.

In field theory, the topological charge is given as~\cite{Polyakov1975} 
\begin{align}
    Q= \frac{1}{4\pi}\int \mathrm{d}^2r\,\mathbf{s}\cdot\left(\partial_x\mathbf{s}\times \partial_y \mathbf{s}\right)\,,\label{eq:Qfield}
\end{align}
where $\mathbf{s}$ is a vector field normalized to unit length. For  lattice spin models this quantity can be calculated as~\cite{Berg1981}:
\begin{align}
     Q=&\sum\limits_{i} q_i\,, \label{eq:Qlattice}
\end{align}
     with the discretized charge density,
\begin{align}
    q_i=&\sum\limits_{\{j,k\}} \frac{1}{2\pi}\mathrm{arctan} \left(\frac{\mathbf{S}_i\cdot (\mathbf{S}_j\times\mathbf{S}_k)} {1 + \mathbf{S}_i\cdot \mathbf{S}_j + \mathbf{S}_i\cdot \mathbf{S}_k + \mathbf{S}_j\cdot \mathbf{S}_k}\right)  \, ,
\end{align}
where $\{\mathbf{S}_i, \mathbf{S}_j, \mathbf{S}_k \}$ denote the spin vectors at three nearest-neighbor sites forming triangles that cover the lattice. In the case of periodic boundary conditions, the value of $Q$ is always an integer, and with this sign convention the topological charge of an isolated skyrmion  in a background pointing along the positive $z$ direction  is $Q=-1$. 

 In contrast to the topological charge, the collective variables in metadynamics are typically continuous. In Ref.~\cite{charalampidis2023arxiv}, Eq.~\eqref{eq:Qfield} was calculated by a numerical discretization of the derivatives, allowing for non-integer topological charges. Because of the focus of that work on single skyrmions, the topological charge was also restricted to low values by the boundary conditions. Here, we calculate the integer topological charge based on Eq.~\eqref{eq:Qlattice}, but also allow for configurations with very densely packed skyrmions, and use the topological charge normalized by the parameter $Q_\mathrm{max}$ to the interval of $(-1,1)$ as the collective variable. We set the normalization to $Q_\mathrm{max}=50$ because at the considered lattice site this value gave energetically unfavourable, very dense stackings of skyrmions. 
 The application of the metadynamics algorithm to atomistic spin models is described in detail in Ref.~\cite{Nagyfalusi2019}.  Following the notations used there, the height of the Gaussian bias potentials was set to $w_0 = 0.04\,\mathrm{mRy}$, with a width of $\sigma = 0.02$, and the metadynamics temperature was set to $T_\mathrm{m} = 1570\,\mathrm{K}$.
The simulations were performed on a two-dimensional triangular lattice of  $128 \times 128$ sites with periodic boundary conditions. 
A multiple-walker algorithm was applied where simulations on five independent replicas 
{contributed to the same bias potential}. Before starting to evolve the bias potential, all the walkers were thermalized
from a collinear field-polarized starting configuration.
The application of multiple walkers ensured the effective mapping of the whole configuration space. The individual bias potentials belonging to each walker were updated after every 25th Monte Carlo step (MCS) with a Gaussian, and they  were  averaged out after every 250th MCS. 
Usually a simulation included $10^6$ MCSs, while this number was increased to $2-5\cdot10^6$ MCSs to obtain better statistics at lower temperatures.

The simulations were performed up to  temperatures where short-range order is completely lost in the systems, and average topological charge approaches zero in a wide external field range.  
However, at low temperatures we encountered severe difficulties  in the simulations  due to the high energy barrier separating the different metastable skyrmionic configurations. If the energy of thermal fluctuations did not overcome this barrier, the time of the simulations necessary to walk through a broad range of the collective variable became extremely long. In our case we were able to go just below  the ordering temperature of the spin-spiral (SS) or skyrmion-lattice (SkL) phases reported in Refs.~\cite{Rozsa2016,Schick2021}. 

In well-tempered metadynamics simulations, the converged value of the bias potential is associated with the free-energy of the system constrained to a given value of the collective variable $Q$ \cite{Dama2014}: 
\begin{align}
    F(Q)=- \frac{T+T_\mathrm{m}}{T_\mathrm{m}}V_\mathrm{bias}(Q)\,.  
\end{align}
At a given temperature, the position of the maximum of $V_\mathrm{bias}(Q)$ (or minimum of $F(Q)$) gives the 
most likely value $Q_0$ of the topological charge in the lattice. 
The average value of the topological charge $\langle Q\rangle$ can be calculated as
\begin{align}
    \langle Q\rangle = \frac {\sum_Q Q\,\mathrm{exp}\{\beta V_\mathrm{bias}(Q)\}} {\sum_Q\mathrm{exp}\{\beta V_\mathrm{bias}(Q)\}}\,,
\end{align}
where $\beta=1/(k_\mathrm{B}T)$ is the inverse temperature.

\section{Results}
\label{sec:results}
 
\subsection{Free-energy curves and chemical potentials}
\label{sec:freeenergy}

 In order to draw conclusions on the thermodynamic properties of skyrmions as quasiparticles, we had to establish the connection between the topological charge used as a collective variable in the system and the number of localized non-collinear spin configurations in the system. Objects with a finite topological charge are not necessarily localized, an elongated segment of a spin spiral can be described by the same topological charge as a skyrmion. In contrast, localized objects with zero net topological charge may also be stable~\cite{Rozsa2017}, which cannot be distinguished from a collinear state based on the collective variable. A value of $Q=1$ may describe an antiskyrmion with the spins winding oppositely compared to a skyrmion, or a skyrmion where all spin directions are reversed compared to the $Q=-1$ case. A value of $Q=-2$ may indicate two individual skyrmions or a composite object. 

   \begin{figure}[htb!] 
   \centering
   \includegraphics[width=1.0\columnwidth] 
 {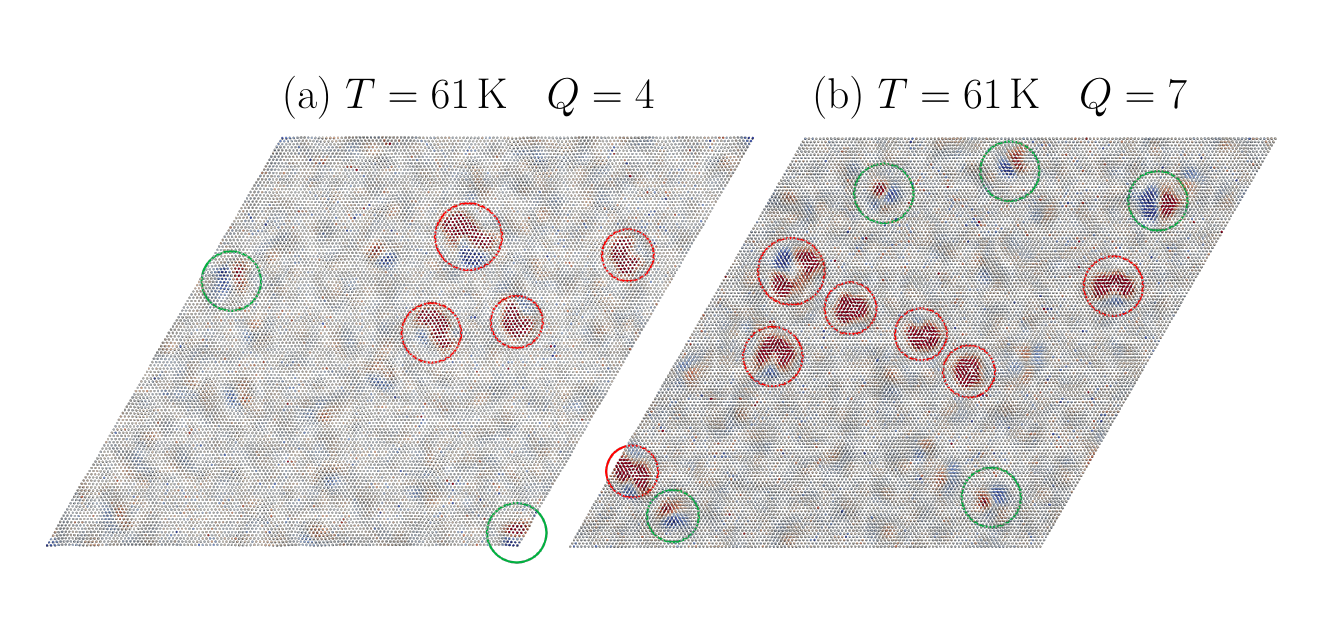} \vspace{-1cm}
 
   \includegraphics[width=1.0\columnwidth]{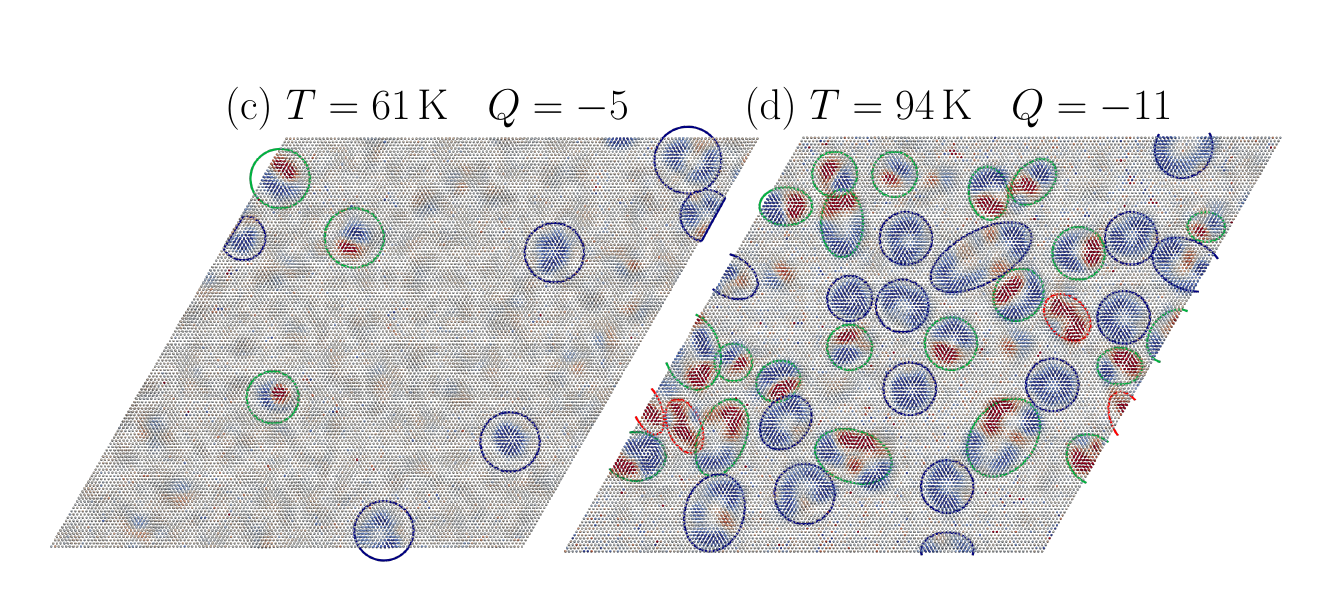} 
   \caption{\label{fig:configurations}  Images of spin configurations  obtained  in (Pt$_{0.95}$Ir$_{0.05}$)/Fe/Pd(111) at $B=1$~T,  with topological charges 
   (a) $Q=4$ at $T$=61\,K, 
   (b) $Q=7$ at $T$=61\,K,
   (c) $Q=-5$ at $T$=61\,K, and
   (d) $Q=-11$ at $T$=94\,K.
    The configurations were relaxed by turning off the thermal fluctuations for better visualization of the localized spin structures; the relaxation did not change the topological charge.
    The coloring denotes the topological charge density: red $q>0$, blue $q<0$. Skyrmionic objects with topological charge $Q=1$, $Q=0$, and $Q=-1$ are enclosed by solid red, black and turquoise lines, respectively. 
    }
  \label{fig:configurations3}
 \end{figure}

 These questions can be qualitatively investigated by looking at real-space spin configurations taken from the simulations.  In  Fig.~\ref{fig:configurations} the spin configurations of Fe sites are displayed for $Q=4$, $Q=7$ and $Q=-5$ at $T=61$\,K and for $Q=-11$  at $T=94$\,K  in (Pt$_{0.95}$Ir$_{0.05}$)/Fe/Pd(111) at $B=1$~T, where skyrmionic objects with many different values of the topological charge are stable at zero temperature~\cite{Rozsa2017}.  The system has a field-polarized ground state at this temperature, but individual skyrmions are created by thermal fluctuations~\cite{Schick2021}. The formation of skyrmions becomes thermodynamically favorable around $T\approx 80~K$.  
For a better identification  of localized objects with a finite topological charge density, 
we relaxed the spin-configuration at zero temperature using an optimized conjugate gradient method~\cite{Nagyfalusi_2022},  thereby smoothening the small-amplitude fluctuations around the collinear state which is not related to skyrmionic structures. We used a clustering algorithm based on the out-of-plane components of the spins to find individual objects with a core pointing opposite to the external magnetic field in the configuration.  Obviously, such an optimisation procedure can alter the topological charge of the system by removing high-energy metastable skyrmionic textures, but especially in the case of low-temperature  simulations  we found that the topological charge was not affected by the optimization. 
We found three types of objects: skyrmions with $Q=-1$ , antiskyrmions with $Q=1$ and chimera skyrmions with $Q=0$ enclosed in turquoise, red and black circles in the figures, respectively.  At this value of the external magnetic field, elongated spiral segments are not present in the simulations, and skyrmions with reversed spin directions, i.e., on a background along the $-z$ direction, are also suppressed. We did not observe composite objects with higher topological charges. At the lower temperature, configurations with a positive net topological charge mostly consist of $Q$ isolated antiskyrmions, as in Fig.~\ref{fig:configurations}(a) and (b), while at positive topological charge $|Q|$ skyrmions may be observed, shown in Fig.~\ref{fig:configurations}(c). Structures with $Q=0$ called chimera skyrmions~\cite{Rozsa2017} can be observed in all configurations which complicate the evaluation of the data, but at the low density of localized objects in Fig.~\ref{fig:configurations}(a)-(c) their effect on the free-energy differences between different $Q$ values is expected to be minor. 

For the temperature $T=94$\,K, the spin configurations in the (Pt$_{0.95}$Ir$_{0.05}$)/Fe/Pd(111) bilayer are only identifiable after the optimisation procedure. Fig.~\ref{fig:configurations}(d) shows the configurations with $Q=-11$, demonstrating
that the topological charge is mostly formed by skyrmions with $Q=-1$, but many other irregularly shaped skyrmion-like patterns with $Q=0$, and also antiskyrmions with $Q=1$ are present in the system. 
Altogether these configurations are much denser than at low $T$, which means the even with a small overall $Q$ for the whole lattice the skyrmion-skyrmion interactions play a crucial role, and an independent particle picture does not apply.

\begin{figure}[htb!]
  \includegraphics[width=1.00\columnwidth]{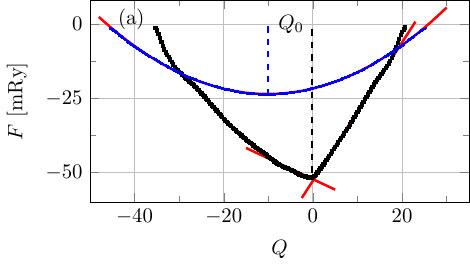}
  \includegraphics[width=1.00\columnwidth]{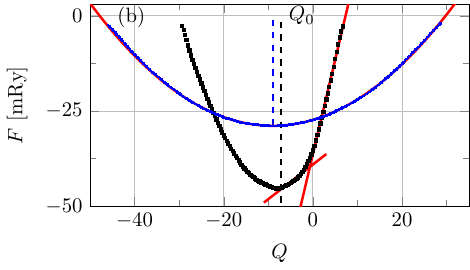}
  \vskip -10pt
  \caption{Free-energy curves as a function of the topological charge $Q$ simulated on a $128\times128$ lattice of (a) Fe sites in (Pt$_{0.95}$Ir$_{0.05}$)/Fe/Pd(111) at $T$=61\,K (black squares) and $T$=94\,K (blue squares) in the presence of an external magnetic field $B=1\,\text{T}$, and (b) in Pd/Fe/Ir(111) at $T$=141\,K (black squares) and $T$=243\,K (blue squares) in the presence of an external magnetic field $B=4\,\text{T}$. $Q_\mathrm{max}$ was set to 50 in both cases. The vertical dashed lines mark the minimum positions $Q_0$ of the free energy $F(Q)$. 
  For the higher temperatures, the solid red lines are obtained by fitting the data to Eq.~\eqref{eq:parabola}, while for the lower temperatures a linear fit was performed in a suitable range below and above $Q=0$.  
  }
  \label{fig:metapot_example}
 \end{figure}

 The real-space configurations can be correlated with the free-energy curves obtained from the metadynamics simulations, shown  in Fig.~\ref{fig:metapot_example}(a)
in case of the (Pt$_{0.95}$Ir$_{0.05}$)/Fe/Pd(111) bilayer in an external field of $B=1\,\mathrm{T}$ for temperatures $T=61\,$K and $T=94$\,K. 
Apparently, $F(Q)$ significantly differs between the two  considered  temperatures. 

\vskip 10pt

The low-temperature free-energy curve has a strongly asymmetric shape with  a minimum $Q_0$ close to zero. 
 The simulation data can be described accurately  in a suitable range above and below $Q=0$ by a linear fit shown by red lines in Fig.~\ref{fig:metapot_example}(a). Based on these linear fits, we defined chemical potentials $\mu_\pm$ of the skyrmionic particles as the slopes of the corresponding curves, 
\begin{align}
    \mu_\pm & =
   \lim\limits_{\Delta Q \to \pm1} \frac{F(\Delta Q)-F(0)}{\left|\Delta Q\right|} 
    \label{eq:lin_fit}
\end{align} 

As can be inferred from  the comparison of Fig.~\ref{fig:configurations}(a) and Fig.~\ref{fig:configurations}(b), in the linear regime above $Q=0$, the increase of the value of the topological charge is related to the creation of antiskyrmions with $Q=1$. Thus, the slope $\mu_+$ of the fitted line in Fig.~\ref{fig:metapot_example}(a) can be identified as the chemical potential of the antiskyrmions. The linearity of $F(Q)$ in the $Q<0$ region persists until $Q=-10$. Similar to the $Q>0$ regime, we can assume  that the dominant process is now the creation of skyrmions, which allows us to identify $\mu_-$ as the chemical potential of the skyrmions. The relations  $\mu_+ > \mu_- > 0$  implies that at $B=1 \, \mathrm{T}$ external field and $T=61 \, \mathrm{K}$ both the skyrmions and antiskyrmions are metastable in (Pt$_{0.95}$Ir$_{0.05}$)/Fe/Pd(111), {{in agreement with the results in Ref.~\cite{Schick2021}.  Proceeding to larger negative $Q$ values, we suppose that more and more skyrmions with $Q=-1$ are created in the system. Beyond a certain density of the skyrmions the interaction between them gives a
 significant  energy contribution, causing a deviation from the linear dependence below about $Q=-10$. 
This implies that the assumption of noninteracting skyrmions persists only in a finite range below $Q_0$. 

At high temperature, a parabola was found to fit well the free-energy curve over the whole range of $Q$ accessed by the simulations, 
\begin{align}
    F(Q)=F_0 + a\,(Q-Q_0)^2\, .
    \label{eq:parabola}
\end{align} 
where $a$ and $Q_0$ are fitting parameters. 
The latter one defines the most likely value of the topological charge, $Q_0 = -10$. 
At this temperature an admixture of different types of skyrmions is being formed as shown in Fig.~\ref{fig:configurations}(d) 
and the change of the free energy with the change of $Q$ cannot be associated with the creation or annihilation of one type of skyrmionic objects.  This explains why a linear fit corresponding to a single type of quasiparticle is not valid at this temperature, leading to the overall parabolic behaviour of the $F(Q)$ curve. 

Figure~\ref{fig:metapot_example}(b) shows the free-energy curves for the Pd/Fe/Ir(111) system at $T =141$\,K  and $T = 243$\,K in the presence of an external magnetic field of $B = 4$\,T,  at which field value the ground state is field-polarized~\cite{Rozsa2016}. The low-temperature spin configurations in Pd/Fe/Ir(111) showed skyrmionic objects with similar behavior as in (Pt$_{0.95}$Ir$_{0.05}$)/Fe/Pd(111). The high-temperature curve is very similar to that of the other bilayer system, showing a parabolic dependence on the topological charge with a value of $Q_0 \simeq -18$. In contrast to the (Pt$_{0.95}$Ir$_{0.05}$)/Fe/Pd(111) system, the low temperature curve is  less asymmetric 
and has a minimum at a $Q_0$ close to that at the higher temperature. Nevertheless, a kink in the curve can be seen around $Q=0$.  Linear fits were performed above and below $Q=0$. Following the definition in Eq.~\eqref{eq:lin_fit}, we obtain $\mu_+>0$ and $\mu_-<0$.  This means that lowering the topological charge, i.e., creating a skyrmion, decreases the free energy of the system.  In the same temperature and field regime, a high number of skyrmions was found in equilibrium in Ref.~\cite{Rozsa2016}, in agreement with this conclusion and the negative value of $Q_{0}$ at the minimum of the free-energy curve.  The relation $\left|\mu_+\right| > \left|\mu_-\right|$ indicates that the free-energy cost of creating an antiskyrmion is larger than the free-energy gain of creating a skyrmion.  Stable antiskyrmions have not been observed at zero temperature in this system in Ref.~\cite{Rozsa2016}, which correlates with the observation that short-lived antiskyrmions created by thermal fluctuations have a high free-energy cost.  
 As it will be discussed in Sec.~\ref{sec:PdFeIr111}, when  decreasing the temperature $|Q_0|$ also decreases, and at a certain temperature $\mu_-(T)$ changes sign,  indicating a field-polarized ground state at lower temperature .  If $Q_0$ is close to zero, the linear regime becomes much narrower,
increasing the numerical error of the fitting procedure.

\subsection{Temperature dependence in (Pt$_{0.95}$Ir$_{0.05}$)/Fe/Pd(111)}
 
 We determined the variation of the chemical potentials defined in Eq.~\eqref{eq:lin_fit} with the temperature.  
First, we performed preliminary calculations in terms of traditional  Metropolis  MC simulations. The heat capacity shown in Fig. \ref{fig:sk_num_Pt95Ir5FePd}(a) set the 
critical temperature around $130\,\mathrm{K}$ which was found to be stable in the range of $B=0.05-1\,\mathrm{T}$. 
The simulated magnetic susceptibility  $\chi\propto\left(\left<\vec{m}^{2}\right>-\left<\vec{m}\right>^{2}\right)/T$, where $\vec{m}$ is the average of the spin vectors over the lattice, 
shows a peak at a lower temperature of about $90\,\mathrm{K}$ in Fig.~\ref{fig:sk_num_Pt95Ir5FePd}(b).  In contrast to a second-order phase transition, the difference between the maxima of the two curves cannot be attributed exclusively to finite-size effects. The susceptibility has a peak where skyrmions and antiskyrmions are created and destroyed very rapidly by the temperature, leading to large fluctuations in the magnetization. In contrast, the heat capacity is maximal where short-range order is completely lost in the system and localized objects may no longer be identified, where the rapidly changing angles between neighbouring spins give rise to high energy fluctuations. 

\begin{figure} 
  \hspace{-0.1cm}
  \includegraphics[width=1.00\columnwidth]{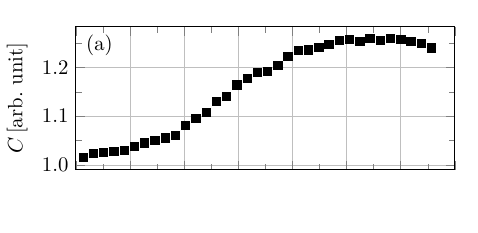} \vspace{-1.7cm}\\
  \includegraphics[width=1\columnwidth]{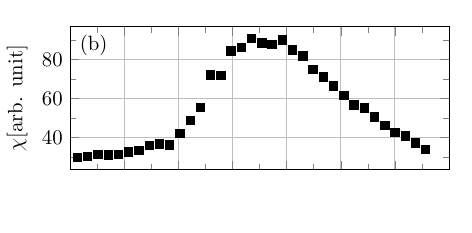} \vspace{-1.7cm}\\
  \includegraphics[width=1\columnwidth]{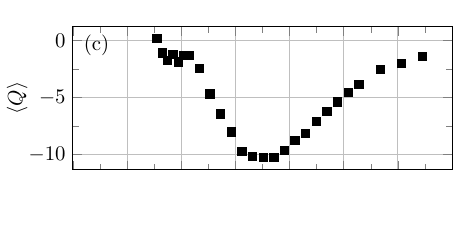} \vspace{-1.7cm}\\
  \includegraphics[width=1\columnwidth]{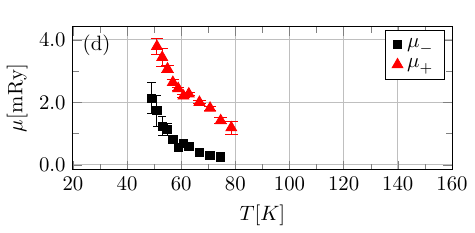}
\caption{ Temperature dependence of thermodynamic parameters in (Pt$_{0.95}$Ir$_{0.05}$)/Fe/Pd(111).  (a) Heat capacity ($C$), 
(b) magnetic spin susceptibility ($\chi$),
(c) average topological charge  $ \langle Q \rangle$, and 
(d) chemical potentials $\mu_\pm$ of skyrmionic quasiparticles 
simulated on a $128\times128$ lattice
with an external magnetic field of $B=1\,$T. $C$ and $\chi$ were obtained from a traditional Monte Carlo simulation, while for the calculation $ \langle Q \rangle$ and $\mu_\pm$ we used metadynamics simulations.
 The error bars  of the chemical potentials  display the uncertainty  of the fitted parameters. 
 }
  \label{fig:sk_num_Pt95Ir5FePd}
 \end{figure}
 
Next we investigated the topological charge of the system by using our metadynamics Monte Carlo method. 
The temperature range of the metadynamics simulations was set between 50\,K and 160\,K. Below $T=48$\,K 
the number of skyrmions did not change even for extremely long simulation times,  and the free-energy curve could not be calculated.  
We found that the average number of skyrmions $\langle Q\rangle$ presented in Fig.~\ref{fig:sk_num_Pt95Ir5FePd}(c)   and its equilibrium value $Q_0$ based on the minimum of the $F(Q)$ curve coincide well with each other.
The shape of the $\langle Q(T) \rangle$ curve is similar to the one reported in Ref.~\cite{Schick2021}, but instead of 
$T\approx 75\, \mathrm{K}$, our method puts the minimum of the curve at a higher temperature of about $90\, \mathrm{K}$.  This may be a consequence of the small system size $25\times25$ used in the simulations in Ref.~\cite{Schick2021}.  
Considering this difference, our result is in good agreement with Ref.~\cite{Schick2021}. 
Apparently, $\langle Q(T) \rangle$ and $\chi(T)$ have their extrema at around the same temperature.
 
The obtained free-energy costs of creating skyrmions and antiskyrmions for the same simulation parameters are shown in Fig.~\ref{fig:sk_num_Pt95Ir5FePd}(d). 
As the chemical potentials could only be defined at low temperatures, where around $Q=0$ the topological charge of the system is dominated by one type of particles, 
the results are limited to a smaller range between $50\,$K and $80\,$K. Above $75-80\,$K the parabolic behavior  illustrated in Fig.~\ref{fig:metapot_example}(a) becomes more pronounced  
in the $F(Q)$ curves. As discussed in Sec.~\ref{sec:freeenergy}, this effect decreases in particular the accuracy of  the fitting of  $\mu_-$. 
Interestingly, the upper boundary of the range where the chemical potentials can be determined coincides with the temperature where the topological charge reaches its minimum.
In the whole temperature range $50 - 80\,$K, $\mu^\mathrm{}_+$ is larger than $\mu^\mathrm{}_-$, i.e., the cost of creating an antiskyrmion is higher than that of creating a skyrmion. 

As can be seen in Fig.~\ref{fig:sk_num_Pt95Ir5FePd}(d), the error bars increase as $T$ decreases due to the less smooth free-energy curves as a consequence of the significantly slower convergence of the metadynamics MC simulations, see Sec.~\ref{sec:methods}. The error also increases near 80\,K, which is explained by the fact that at this temperature secondary particles can appear more easily, making the fitting  less accurate.  

\subsection{Temperature dependence in Pd/Fe/Ir(111)}
\label{sec:PdFeIr111}

 We also calculated the thermodynamic quantities in the Pd/Fe/Ir(111) system, where the skyrmion lattice phase is stable at zero temperature.  As for the (Pt$_{0.95}$Ir$_{0.05}$)/Fe/Pd(111) bilayer, first we determined 
the heat capacity  from Metropolis MC simulations .
 We found that the heat capacity has a peak  around 250\,K-280\,K depending on the magnitude of the external field.  This value is close to the maximum of the heat capacity simulated in Ref.~\cite{Bottcher2018}, where the same system was investigated based on a differently parametrized spin model, and also agree with the temperature regime where short-range order was found to vanish in Ref.~\cite{Rozsa2016} using the same parameters as in the present work.  
Based on this result, we set the upper limit of the metadynamics simulations to {370\,K}. 
Corresponding to the different ground states  of the system~\cite{Rozsa2016}, in our simulations we choose the following three external field values: $B=0.67\,\mathrm{T}$ with a spin-spiral ground state,  $B=1.51\,\mathrm{T}$ with a skyrmion lattice ground state, and  $B=4\,\mathrm{T}$, where the system is expected to have a field-polarized ground state.
Similar to (Pt$_{0.95}$Ir$_{0.05}$)/Fe/Pd(111), the equilibrium and the most likely value of the topological charge differ only a little, therefore,
in Fig. \ref{fig:sk_num_FePdIr}(a) we only present the 
most likely value $Q_0$ as a function of the temperature for the chosen values of the external magnetic field. 

\begin{figure}[htb]
  \includegraphics[width=1.00\columnwidth]{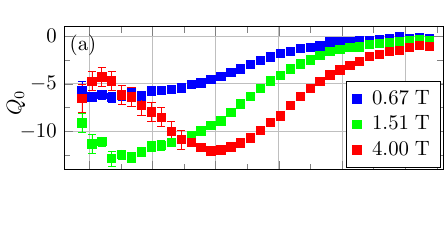}  \vspace{-1.7cm}\\
  \includegraphics[width=1\columnwidth]{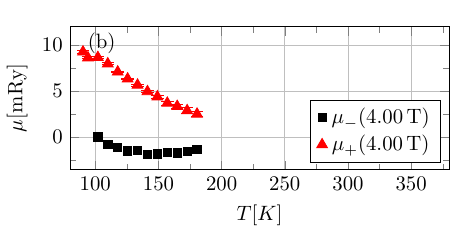}
  \caption{ Temperature dependence of thermodynamic parameters in Pd/Fe/Ir(111).  (a) 
  Most likely  value of the  topological charge simulated on a $128\times128$ lattice 
  in the presence of different external magnetic fields, $0.67\,$T, $1.51\,$T  and $4.00\,$T. (b) The chemical potential of skyrmions ($-$) and antiskyrmions ($+$) on a $128\times128$ lattice of Pd/Fe/Ir(111) surface with  external magnetic field $B=4.00\,$T. The error bars display the uncertainty  of the fitting parameters. 
  }
  \label{fig:sk_num_FePdIr}
 \end{figure}

It can be seen from Fig. \ref{fig:sk_num_FePdIr}(a) that the three curves show a maximum of the skyrmion number $|Q_0|$  at intermediate temperatures.  
 This indicates that the formation of skyrmions becomes energetically more favorable when the temperature is increased, similarly to the (Pt$_{0.95}$Ir$_{0.05}$)/Fe/Pd(111) bilayer. The temperature dependence of the skyrmion number agrees well with the Metropolis MC simulations in Ref.~\cite{Rozsa2016}. 

The investigation of the corresponding spin configurations indicated that in case of SkL ground state ($B=4$~T) at low temperatures the topological charge exclusively corresponds to skyrmions with $Q=-1$. At higher temperatures minority particles such as antiskyrmions ($Q=1$) and chimera skyrmions ($Q=0$) are also present. 
 At very high temperature well-defined localized configurations can no longer be identified, and $|Q_0|$ converges to zero since the topological charge cannot be meaningfully used to characterize the spin structure in this regime. 

For $B=1.51\,\mathrm{T}$, the maximum  value  of the skyrmion number is similar to that for $B=4\,\mathrm{T}$, but it is reached at a considerably lower temperature ($\approx$\,120\,K) and the decrease of $|Q_0|$ towards $T=0$ is less pronounced. 
 This is expected since the ground state should be a skyrmion lattice at this field value~\cite{Rozsa2016}.  
For $B=0.67\,\mathrm{T}$, the maximum number of skyrmions is the smallest among the three cases, and it hardly decreases for temperatures below the maximum.  Theoretically, the topological charge should vanish in the zero-temperature limit, because the spin-spiral state is the ground state in this case. The real-space spin configurations at $B=0.67$~T and $B=1.51$~T resemble space-filling networks of domain walls in the low-temperature regime. Individual skyrmions are not stable at zero temperature at these low field values, instead they fill up the whole space due to their strip-out instability. The topological charge density is enhanced at the ends of the intertwined spin-spiral segments, but localized configurations with an integer topological charge can no longer observed. Due to the strong interaction between these elongated objects, the independent-particle picture underlying Eq.~\eqref{eq:lin_fit} is no longer valid, and the free-energy curves are parabolic. Therefore, the chemical potentials were only calculated for $B=4.00$~T.

The temperature dependence of the chemical potentials $\mu_\pm$ is shown in Fig.~\ref{fig:sk_num_FePdIr}(b). 
For $B=4\, \mathrm{T}$ where the ground state configuration is the field-polarized state, the linear fitting around $Q=0$ was possible up to the temperature where the 
most likely  value of the  topological charge has its minimum at about 180\,K.
As can be seen in Fig.~\ref{fig:sk_num_FePdIr}(b), the free-energy cost for the creation of antiskyrmions $\mu_+$ is positive, while that of the skyrmions $\mu_-$ is negative, as was shown in Sec.~\ref{sec:freeenergy}. 
This implies that in this temperature region the formation of skyrmions with $Q=-1$ is favorable, whereas the formation of antiskyrmions with $Q=1$ is highly unfavorable.  As the most likely value of the topological charge approaches zero at lower temperature, the linear regime for $Q<0$ contracts. This makes it difficult to determine the chemical potential based on the fitting procedure when the slope is close to zero. This is the reason why $\mu_{+}$ could be determined at a few temperature values below $100$~K in Fig.~\ref{fig:sk_num_FePdIr}(b), but it was not possible to calculate $\mu_{-}$ from the same simulations. It is expected that the chemical potential changes sign in this regime, leading to the field-polarized ground state at zero temperature.

 \section{Conclusion}
 \label{sec:conclusion}
 
We evaluated the free-energy surface of (Pt$_{0.95}$Ir$_{0.05}$)/Fe/Pd(111) and Pd/Fe/Ir(111) bilayers as a function of the topological charge in terms of metadynamics Monte Carlo simulations by employing first-principles-based spin models. 
 We found that the most likely value of the topological charge shifts away from zero with increasing temperature for field values where the ground-state is spin-polarized, indicating that thermal fluctuations favor skyrmion formation, in agreement with previous results on both systems~\cite{Rozsa2016,Bottcher2018,Schick2021}. 
Beyond determining the equilibrium value of the skyrmion quasiparticles, the knowledge of the whole free-energy surface made it possible to extract the free-energy cost of creating a skyrmion or an antiskyrmion 
 from the topologically trivial configuration by a linear fit to the free-energy function , which we identified as the chemical potential of these quasiparticles. The two kinds of chemical potentials differ from each other: for the investigated systems the creation of antiskyrmions costs more free energy than that of a skyrmion.  We found that this method for calculating the chemical potential is only valid if the skyrmions can be treated as independent quasiparticles, which assumption breaks down at high temperature where short-range order is completely lost, at low field values where the skyrmions elongate into spin spiral segments and fill up the whole space, or when the skyrmion density is high such as in the skyrmion lattice phase. In these cases the interactions between the localized configurations manifests itself in a parabolic free-energy curve in the vicinity of the minimum. Future investigations of the thermodynamic properties of interacting skyrmions could shed further light on the phase formation of these quasiparticles.  

\section*{Acknowledgements}
This research was supported by the National Research, Development, and Innovation (NRDI) Office of Hungary under Project Nos. K131938, K142652 and FK142601. Additional support was provided by the Ministry of Culture and Innovation and the NRDI Office within the Quantum Information National Laboratory of Hungary (Grant No. 2022-2.1.1-NL-2022-00004), and by the Hungarian Academy of Sciences via a János Bolyai Research Grant (Grant No. BO/00178/23/11). This project has received funding from the HUN-REN Hungarian Research Network.

\bibliography{skyrmioncikk}{}

\begin{thebibliography}{50}%
\makeatletter
\providecommand \@ifxundefined [1]{%
 \@ifx{#1\undefined}
}%
\providecommand \@ifnum [1]{%
 \ifnum #1\expandafter \@firstoftwo
 \else \expandafter \@secondoftwo
 \fi
}%
\providecommand \@ifx [1]{%
 \ifx #1\expandafter \@firstoftwo
 \else \expandafter \@secondoftwo
 \fi
}%
\providecommand \natexlab [1]{#1}%
\providecommand \enquote  [1]{``#1''}%
\providecommand \bibnamefont  [1]{#1}%
\providecommand \bibfnamefont [1]{#1}%
\providecommand \citenamefont [1]{#1}%
\providecommand \href@noop [0]{\@secondoftwo}%
\providecommand \href [0]{\begingroup \@sanitize@url \@href}%
\providecommand \@href[1]{\@@startlink{#1}\@@href}%
\providecommand \@@href[1]{\endgroup#1\@@endlink}%
\providecommand \@sanitize@url [0]{\catcode `\\12\catcode `\$12\catcode
  `\&12\catcode `\#12\catcode `\^12\catcode `\_12\catcode `\%12\relax}%
\providecommand \@@startlink[1]{}%
\providecommand \@@endlink[0]{}%
\providecommand \url  [0]{\begingroup\@sanitize@url \@url }%
\providecommand \@url [1]{\endgroup\@href {#1}{\urlprefix }}%
\providecommand \urlprefix  [0]{URL }%
\providecommand \Eprint [0]{\href }%
\providecommand \doibase [0]{https://doi.org/}%
\providecommand \selectlanguage [0]{\@gobble}%
\providecommand \bibinfo  [0]{\@secondoftwo}%
\providecommand \bibfield  [0]{\@secondoftwo}%
\providecommand \translation [1]{[#1]}%
\providecommand \BibitemOpen [0]{}%
\providecommand \bibitemStop [0]{}%
\providecommand \bibitemNoStop [0]{.\EOS\space}%
\providecommand \EOS [0]{\spacefactor3000\relax}%
\providecommand \BibitemShut  [1]{\csname bibitem#1\endcsname}%
\let\auto@bib@innerbib\@empty
\bibitem [{\citenamefont {Mühlbauer}\ \emph {et~al.}(2009)\citenamefont
  {Mühlbauer}, \citenamefont {Binz}, \citenamefont {Jonietz}, \citenamefont
  {Pfleiderer}, \citenamefont {Rosch}, \citenamefont {Neubauer}, \citenamefont
  {Georgii},\ and\ \citenamefont {Böni}}]{Muhlbauer2009}%
  \BibitemOpen
  \bibfield  {author} {\bibinfo {author} {\bibfnamefont {S.}~\bibnamefont
  {Mühlbauer}}, \bibinfo {author} {\bibfnamefont {B.}~\bibnamefont {Binz}},
  \bibinfo {author} {\bibfnamefont {F.}~\bibnamefont {Jonietz}}, \bibinfo
  {author} {\bibfnamefont {C.}~\bibnamefont {Pfleiderer}}, \bibinfo {author}
  {\bibfnamefont {A.}~\bibnamefont {Rosch}}, \bibinfo {author} {\bibfnamefont
  {A.}~\bibnamefont {Neubauer}}, \bibinfo {author} {\bibfnamefont
  {R.}~\bibnamefont {Georgii}},\ and\ \bibinfo {author} {\bibfnamefont
  {P.}~\bibnamefont {Böni}},\ }\bibfield  {title} {\bibinfo {title} {{Skyrmion
  Lattice in a Chiral Magnet}},\ }\href
  {https://doi.org/10.1126/science.1166767} {\bibfield  {journal} {\bibinfo
  {journal} {Science}\ }\textbf {\bibinfo {volume} {323}},\ \bibinfo {pages}
  {915} (\bibinfo {year} {2009})}\BibitemShut {NoStop}%
\bibitem [{\citenamefont {Neubauer}\ \emph {et~al.}(2009)\citenamefont
  {Neubauer}, \citenamefont {Pfleiderer}, \citenamefont {Binz}, \citenamefont
  {Rosch}, \citenamefont {Ritz}, \citenamefont {Niklowitz},\ and\ \citenamefont
  {B\"oni}}]{Neubauer2009}%
  \BibitemOpen
  \bibfield  {author} {\bibinfo {author} {\bibfnamefont {A.}~\bibnamefont
  {Neubauer}}, \bibinfo {author} {\bibfnamefont {C.}~\bibnamefont
  {Pfleiderer}}, \bibinfo {author} {\bibfnamefont {B.}~\bibnamefont {Binz}},
  \bibinfo {author} {\bibfnamefont {A.}~\bibnamefont {Rosch}}, \bibinfo
  {author} {\bibfnamefont {R.}~\bibnamefont {Ritz}}, \bibinfo {author}
  {\bibfnamefont {P.~G.}\ \bibnamefont {Niklowitz}},\ and\ \bibinfo {author}
  {\bibfnamefont {P.}~\bibnamefont {B\"oni}},\ }\bibfield  {title} {\bibinfo
  {title} {{Topological Hall Effect in the $A$ Phase of MnSi}},\ }\href
  {https://doi.org/10.1103/PhysRevLett.102.186602} {\bibfield  {journal}
  {\bibinfo  {journal} {Phys. Rev. Lett.}\ }\textbf {\bibinfo {volume} {102}},\
  \bibinfo {pages} {186602} (\bibinfo {year} {2009})}\BibitemShut {NoStop}%
\bibitem [{\citenamefont {Yu}\ \emph {et~al.}(2010)\citenamefont {Yu},
  \citenamefont {Onose}, \citenamefont {Kanazawa}, \citenamefont {Park},
  \citenamefont {Han}, \citenamefont {Matsui}, \citenamefont {Nagaosa},\ and\
  \citenamefont {Tokura}}]{Yu2010}%
  \BibitemOpen
  \bibfield  {author} {\bibinfo {author} {\bibfnamefont {X.~Z.}\ \bibnamefont
  {Yu}}, \bibinfo {author} {\bibfnamefont {Y.}~\bibnamefont {Onose}}, \bibinfo
  {author} {\bibfnamefont {N.}~\bibnamefont {Kanazawa}}, \bibinfo {author}
  {\bibfnamefont {J.~H.}\ \bibnamefont {Park}}, \bibinfo {author}
  {\bibfnamefont {J.~H.}\ \bibnamefont {Han}}, \bibinfo {author} {\bibfnamefont
  {Y.}~\bibnamefont {Matsui}}, \bibinfo {author} {\bibfnamefont
  {N.}~\bibnamefont {Nagaosa}},\ and\ \bibinfo {author} {\bibfnamefont
  {Y.}~\bibnamefont {Tokura}},\ }\bibfield  {title} {\bibinfo {title}
  {{Real-space observation of a two-dimensional skyrmion crystal}},\ }\href
  {https://doi.org/10.1038/nature09124} {\bibfield  {journal} {\bibinfo
  {journal} {Nature}\ }\textbf {\bibinfo {volume} {465}},\ \bibinfo {pages}
  {901} (\bibinfo {year} {2010})}\BibitemShut {NoStop}%
\bibitem [{\citenamefont {Tokura}\ and\ \citenamefont
  {Kanazawa}(2021)}]{Tokura2021}%
  \BibitemOpen
  \bibfield  {author} {\bibinfo {author} {\bibfnamefont {Y.}~\bibnamefont
  {Tokura}}\ and\ \bibinfo {author} {\bibfnamefont {N.}~\bibnamefont
  {Kanazawa}},\ }\bibfield  {title} {\bibinfo {title} {{Magnetic Skyrmion
  Materials}},\ }\href {https://doi.org/10.1021/acs.chemrev.0c00297} {\bibfield
   {journal} {\bibinfo  {journal} {Chemical Reviews}\ }\textbf {\bibinfo
  {volume} {121}},\ \bibinfo {pages} {2857} (\bibinfo {year} {2021})},\
  \bibinfo {note} {pMID: 33164494}\BibitemShut {NoStop}%
\bibitem [{\citenamefont {Back}\ \emph {et~al.}(2020)\citenamefont {Back},
  \citenamefont {Cros}, \citenamefont {Ebert}, \citenamefont {Everschor-Sitte},
  \citenamefont {Fert}, \citenamefont {Garst}, \citenamefont {Ma},
  \citenamefont {Mankovsky}, \citenamefont {Monchesky}, \citenamefont
  {Mostovoy}, \citenamefont {Nagaosa}, \citenamefont {Parkin}, \citenamefont
  {Pfleiderer}, \citenamefont {Reyren}, \citenamefont {Rosch}, \citenamefont
  {Taguchi}, \citenamefont {Tokura}, \citenamefont {von Bergmann},\ and\
  \citenamefont {Zang}}]{Back_2020}%
  \BibitemOpen
  \bibfield  {author} {\bibinfo {author} {\bibfnamefont {C.}~\bibnamefont
  {Back}}, \bibinfo {author} {\bibfnamefont {V.}~\bibnamefont {Cros}}, \bibinfo
  {author} {\bibfnamefont {H.}~\bibnamefont {Ebert}}, \bibinfo {author}
  {\bibfnamefont {K.}~\bibnamefont {Everschor-Sitte}}, \bibinfo {author}
  {\bibfnamefont {A.}~\bibnamefont {Fert}}, \bibinfo {author} {\bibfnamefont
  {M.}~\bibnamefont {Garst}}, \bibinfo {author} {\bibfnamefont
  {T.}~\bibnamefont {Ma}}, \bibinfo {author} {\bibfnamefont {S.}~\bibnamefont
  {Mankovsky}}, \bibinfo {author} {\bibfnamefont {T.~L.}\ \bibnamefont
  {Monchesky}}, \bibinfo {author} {\bibfnamefont {M.}~\bibnamefont {Mostovoy}},
  \bibinfo {author} {\bibfnamefont {N.}~\bibnamefont {Nagaosa}}, \bibinfo
  {author} {\bibfnamefont {S.~S.~P.}\ \bibnamefont {Parkin}}, \bibinfo {author}
  {\bibfnamefont {C.}~\bibnamefont {Pfleiderer}}, \bibinfo {author}
  {\bibfnamefont {N.}~\bibnamefont {Reyren}}, \bibinfo {author} {\bibfnamefont
  {A.}~\bibnamefont {Rosch}}, \bibinfo {author} {\bibfnamefont
  {Y.}~\bibnamefont {Taguchi}}, \bibinfo {author} {\bibfnamefont
  {Y.}~\bibnamefont {Tokura}}, \bibinfo {author} {\bibfnamefont
  {K.}~\bibnamefont {von Bergmann}},\ and\ \bibinfo {author} {\bibfnamefont
  {J.}~\bibnamefont {Zang}},\ }\bibfield  {title} {\bibinfo {title} {{The 2020
  skyrmionics roadmap}},\ }\href {https://doi.org/10.1088/1361-6463/ab8418}
  {\bibfield  {journal} {\bibinfo  {journal} {Journal of Physics D: Applied
  Physics}\ }\textbf {\bibinfo {volume} {53}},\ \bibinfo {pages} {363001}
  (\bibinfo {year} {2020})}\BibitemShut {NoStop}%
\bibitem [{\citenamefont {Dzyaloshinsky}(1958)}]{DZYALOSHINSKY1958}%
  \BibitemOpen
  \bibfield  {author} {\bibinfo {author} {\bibfnamefont {I.}~\bibnamefont
  {Dzyaloshinsky}},\ }\bibfield  {title} {\bibinfo {title} {{A thermodynamic
  theory of “weak” ferromagnetism of antiferromagnetics}},\ }\href
  {https://doi.org/https://doi.org/10.1016/0022-3697(58)90076-3} {\bibfield
  {journal} {\bibinfo  {journal} {Journal of Physics and Chemistry of Solids}\
  }\textbf {\bibinfo {volume} {4}},\ \bibinfo {pages} {241} (\bibinfo {year}
  {1958})}\BibitemShut {NoStop}%
\bibitem [{\citenamefont {Moriya}(1960)}]{Moriya1960}%
  \BibitemOpen
  \bibfield  {author} {\bibinfo {author} {\bibfnamefont {T.}~\bibnamefont
  {Moriya}},\ }\bibfield  {title} {\bibinfo {title} {{New Mechanism of
  Anisotropic Superexchange Interaction}},\ }\href
  {https://doi.org/10.1103/PhysRevLett.4.228} {\bibfield  {journal} {\bibinfo
  {journal} {Phys. Rev. Lett.}\ }\textbf {\bibinfo {volume} {4}},\ \bibinfo
  {pages} {228} (\bibinfo {year} {1960})}\BibitemShut {NoStop}%
\bibitem [{\citenamefont {Bogdanov}\ and\ \citenamefont
  {Hubert}(1994)}]{Bogdanov1994}%
  \BibitemOpen
  \bibfield  {author} {\bibinfo {author} {\bibfnamefont {A.}~\bibnamefont
  {Bogdanov}}\ and\ \bibinfo {author} {\bibfnamefont {A.}~\bibnamefont
  {Hubert}},\ }\bibfield  {title} {\bibinfo {title} {{Thermodynamically stable
  magnetic vortex states in magnetic crystals}},\ }\href
  {https://doi.org/https://doi.org/10.1016/0304-8853(94)90046-9} {\bibfield
  {journal} {\bibinfo  {journal} {Journal of Magnetism and Magnetic Materials}\
  }\textbf {\bibinfo {volume} {138}},\ \bibinfo {pages} {255} (\bibinfo {year}
  {1994})}\BibitemShut {NoStop}%
\bibitem [{\citenamefont {Heinze}\ \emph {et~al.}(2011)\citenamefont {Heinze},
  \citenamefont {von Bergmann}, \citenamefont {Menzel}, \citenamefont {Brede},
  \citenamefont {Kubetzka}, \citenamefont {Wiesendanger}, \citenamefont
  {Bihlmayer},\ and\ \citenamefont {Blügel}}]{Heinze2011}%
  \BibitemOpen
  \bibfield  {author} {\bibinfo {author} {\bibfnamefont {S.}~\bibnamefont
  {Heinze}}, \bibinfo {author} {\bibfnamefont {K.}~\bibnamefont {von
  Bergmann}}, \bibinfo {author} {\bibfnamefont {M.}~\bibnamefont {Menzel}},
  \bibinfo {author} {\bibfnamefont {J.}~\bibnamefont {Brede}}, \bibinfo
  {author} {\bibfnamefont {A.}~\bibnamefont {Kubetzka}}, \bibinfo {author}
  {\bibfnamefont {R.}~\bibnamefont {Wiesendanger}}, \bibinfo {author}
  {\bibfnamefont {G.}~\bibnamefont {Bihlmayer}},\ and\ \bibinfo {author}
  {\bibfnamefont {S.}~\bibnamefont {Blügel}},\ }\bibfield  {title} {\bibinfo
  {title} {{Spontaneous atomic-scale magnetic skyrmion lattice in two
  dimensions}},\ }\href {https://doi.org/https://doi.org/10.1038/nphys2045}
  {\bibfield  {journal} {\bibinfo  {journal} {Nature Physics}\ }\textbf
  {\bibinfo {volume} {7}},\ \bibinfo {pages} {713–718} (\bibinfo {year}
  {2011})}\BibitemShut {NoStop}%
\bibitem [{\citenamefont {Okubo}\ \emph {et~al.}(2012)\citenamefont {Okubo},
  \citenamefont {Chung},\ and\ \citenamefont {Kawamura}}]{Okubo2012}%
  \BibitemOpen
  \bibfield  {author} {\bibinfo {author} {\bibfnamefont {T.}~\bibnamefont
  {Okubo}}, \bibinfo {author} {\bibfnamefont {S.}~\bibnamefont {Chung}},\ and\
  \bibinfo {author} {\bibfnamefont {H.}~\bibnamefont {Kawamura}},\ }\bibfield
  {title} {\bibinfo {title} {{Multiple-$q$ States and the Skyrmion Lattice of
  the Triangular-Lattice Heisenberg Antiferromagnet under Magnetic Fields}},\
  }\href {https://doi.org/10.1103/PhysRevLett.108.017206} {\bibfield  {journal}
  {\bibinfo  {journal} {Phys. Rev. Lett.}\ }\textbf {\bibinfo {volume} {108}},\
  \bibinfo {pages} {017206} (\bibinfo {year} {2012})}\BibitemShut {NoStop}%
\bibitem [{\citenamefont {von Malottki}\ \emph {et~al.}(2017)\citenamefont {von
  Malottki}, \citenamefont {Dupé}, \citenamefont {Delin},\ and\ \citenamefont
  {Heinze}}]{vonMalottki2017}%
  \BibitemOpen
  \bibfield  {author} {\bibinfo {author} {\bibfnamefont {S.}~\bibnamefont {von
  Malottki}}, \bibinfo {author} {\bibfnamefont {P.~F.}\ \bibnamefont {Dupé},
  \bibfnamefont {B.~andBessarab}}, \bibinfo {author} {\bibfnamefont
  {A.}~\bibnamefont {Delin}},\ and\ \bibinfo {author} {\bibfnamefont
  {S.}~\bibnamefont {Heinze}},\ }\bibfield  {title} {\bibinfo {title}
  {{Enhanced skyrmion stability due to exchange frustration}},\ }\href
  {https://doi.org/10.1038/s41598-017-12525-x} {\bibfield  {journal} {\bibinfo
  {journal} {Scientific Reports}\ }\textbf {\bibinfo {volume} {7}},\ \bibinfo
  {pages} {12299} (\bibinfo {year} {2017})}\BibitemShut {NoStop}%
\bibitem [{\citenamefont {Fert}\ \emph {et~al.}(2013)\citenamefont {Fert},
  \citenamefont {Cros},\ and\ \citenamefont {Sampaio}}]{Fert2013}%
  \BibitemOpen
  \bibfield  {author} {\bibinfo {author} {\bibfnamefont {A.}~\bibnamefont
  {Fert}}, \bibinfo {author} {\bibfnamefont {V.}~\bibnamefont {Cros}},\ and\
  \bibinfo {author} {\bibfnamefont {J.}~\bibnamefont {Sampaio}},\ }\bibfield
  {title} {\bibinfo {title} {{Skyrmions on the track}},\ }\href
  {https://doi.org/10.1038/nnano.2013.29} {\bibfield  {journal} {\bibinfo
  {journal} {Nature Nanotechnology}\ }\textbf {\bibinfo {volume} {8}},\
  \bibinfo {pages} {152} (\bibinfo {year} {2013})}\BibitemShut {NoStop}%
\bibitem [{\citenamefont {Iwasaki}\ \emph {et~al.}(2013)\citenamefont
  {Iwasaki}, \citenamefont {Mochizuki},\ and\ \citenamefont
  {Nagaosa}}]{Iwasaki2013}%
  \BibitemOpen
  \bibfield  {author} {\bibinfo {author} {\bibfnamefont {J.}~\bibnamefont
  {Iwasaki}}, \bibinfo {author} {\bibfnamefont {M.}~\bibnamefont {Mochizuki}},\
  and\ \bibinfo {author} {\bibfnamefont {N.}~\bibnamefont {Nagaosa}},\
  }\bibfield  {title} {\bibinfo {title} {{Current-induced skyrmion dynamics in
  constricted geometries}},\ }\href {https://doi.org/10.1038/nnano.2013.176}
  {\bibfield  {journal} {\bibinfo  {journal} {Nature Nanotechnology}\ }\textbf
  {\bibinfo {volume} {8}},\ \bibinfo {pages} {742} (\bibinfo {year}
  {2013})}\BibitemShut {NoStop}%
\bibitem [{\citenamefont {Zhou}\ and\ \citenamefont {Ezawa}(2014)}]{Zhou2014}%
  \BibitemOpen
  \bibfield  {author} {\bibinfo {author} {\bibfnamefont {Y.}~\bibnamefont
  {Zhou}}\ and\ \bibinfo {author} {\bibfnamefont {M.}~\bibnamefont {Ezawa}},\
  }\bibfield  {title} {\bibinfo {title} {{A reversible conversion between a
  skyrmion and a domain-wall pair in a junction geometry}},\ }\href
  {https://doi.org/10.1038/ncomms5652} {\bibfield  {journal} {\bibinfo
  {journal} {Nature Communications}\ }\textbf {\bibinfo {volume} {5}},\
  \bibinfo {pages} {4652} (\bibinfo {year} {2014})}\BibitemShut {NoStop}%
\bibitem [{\citenamefont {Grollier}\ \emph {et~al.}(2020)\citenamefont
  {Grollier}, \citenamefont {Querlioz}, \citenamefont {Camsari}, \citenamefont
  {Everschor-Sitte}, \citenamefont {Fukami},\ and\ \citenamefont
  {Stiles}}]{Grollier2020}%
  \BibitemOpen
  \bibfield  {author} {\bibinfo {author} {\bibfnamefont {J.}~\bibnamefont
  {Grollier}}, \bibinfo {author} {\bibfnamefont {D.}~\bibnamefont {Querlioz}},
  \bibinfo {author} {\bibfnamefont {K.~Y.}\ \bibnamefont {Camsari}}, \bibinfo
  {author} {\bibfnamefont {K.}~\bibnamefont {Everschor-Sitte}}, \bibinfo
  {author} {\bibfnamefont {S.}~\bibnamefont {Fukami}},\ and\ \bibinfo {author}
  {\bibfnamefont {M.~D.}\ \bibnamefont {Stiles}},\ }\bibfield  {title}
  {\bibinfo {title} {{Neuromorphic spintronics}},\ }\href
  {https://doi.org/10.1038/s41928-019-0360-9} {\bibfield  {journal} {\bibinfo
  {journal} {Nature Electronics}\ }\textbf {\bibinfo {volume} {3}},\ \bibinfo
  {pages} {360} (\bibinfo {year} {2020})}\BibitemShut {NoStop}%
\bibitem [{\citenamefont {Psaroudaki}\ and\ \citenamefont
  {Panagopoulos}(2021)}]{Psaroudaki2021}%
  \BibitemOpen
  \bibfield  {author} {\bibinfo {author} {\bibfnamefont {C.}~\bibnamefont
  {Psaroudaki}}\ and\ \bibinfo {author} {\bibfnamefont {C.}~\bibnamefont
  {Panagopoulos}},\ }\bibfield  {title} {\bibinfo {title} {{Skyrmion Qubits: A
  New Class of Quantum Logic Elements Based on Nanoscale Magnetization}},\
  }\href {https://doi.org/10.1103/PhysRevLett.127.067201} {\bibfield  {journal}
  {\bibinfo  {journal} {Phys. Rev. Lett.}\ }\textbf {\bibinfo {volume} {127}},\
  \bibinfo {pages} {067201} (\bibinfo {year} {2021})}\BibitemShut {NoStop}%
\bibitem [{\citenamefont {Romming}\ \emph {et~al.}(2013)\citenamefont
  {Romming}, \citenamefont {Hanneken}, \citenamefont {Menzel}, \citenamefont
  {Bickel}, \citenamefont {Wolter}, \citenamefont {von Bergmann}, \citenamefont
  {Kubetzka},\ and\ \citenamefont {Wiesendanger}}]{Romming2013}%
  \BibitemOpen
  \bibfield  {author} {\bibinfo {author} {\bibfnamefont {N.}~\bibnamefont
  {Romming}}, \bibinfo {author} {\bibfnamefont {C.}~\bibnamefont {Hanneken}},
  \bibinfo {author} {\bibfnamefont {M.}~\bibnamefont {Menzel}}, \bibinfo
  {author} {\bibfnamefont {J.~E.}\ \bibnamefont {Bickel}}, \bibinfo {author}
  {\bibfnamefont {B.}~\bibnamefont {Wolter}}, \bibinfo {author} {\bibfnamefont
  {K.}~\bibnamefont {von Bergmann}}, \bibinfo {author} {\bibfnamefont
  {A.}~\bibnamefont {Kubetzka}},\ and\ \bibinfo {author} {\bibfnamefont
  {R.}~\bibnamefont {Wiesendanger}},\ }\bibfield  {title} {\bibinfo {title}
  {{Writing and Deleting Single Magnetic Skyrmions}},\ }\href
  {https://doi.org/10.1126/science.1240573} {\bibfield  {journal} {\bibinfo
  {journal} {Science}\ }\textbf {\bibinfo {volume} {341}},\ \bibinfo {pages}
  {636} (\bibinfo {year} {2013})}\BibitemShut {NoStop}%
\bibitem [{\citenamefont {Romming}\ \emph {et~al.}(2015)\citenamefont
  {Romming}, \citenamefont {Kubetzka}, \citenamefont {Hanneken}, \citenamefont
  {von Bergmann},\ and\ \citenamefont {Wiesendanger}}]{Romming2015}%
  \BibitemOpen
  \bibfield  {author} {\bibinfo {author} {\bibfnamefont {N.}~\bibnamefont
  {Romming}}, \bibinfo {author} {\bibfnamefont {A.}~\bibnamefont {Kubetzka}},
  \bibinfo {author} {\bibfnamefont {C.}~\bibnamefont {Hanneken}}, \bibinfo
  {author} {\bibfnamefont {K.}~\bibnamefont {von Bergmann}},\ and\ \bibinfo
  {author} {\bibfnamefont {R.}~\bibnamefont {Wiesendanger}},\ }\bibfield
  {title} {\bibinfo {title} {{Field-Dependent Size and Shape of Single Magnetic
  Skyrmions}},\ }\href {https://doi.org/10.1103/PhysRevLett.114.177203}
  {\bibfield  {journal} {\bibinfo  {journal} {Phys. Rev. Lett.}\ }\textbf
  {\bibinfo {volume} {114}},\ \bibinfo {pages} {177203} (\bibinfo {year}
  {2015})}\BibitemShut {NoStop}%
\bibitem [{\citenamefont {R\'ozsa}\ \emph
  {et~al.}(2016{\natexlab{a}})\citenamefont {R\'ozsa}, \citenamefont {Simon},
  \citenamefont {Palot\'as}, \citenamefont {Udvardi},\ and\ \citenamefont
  {Szunyogh}}]{Rozsa2016}%
  \BibitemOpen
  \bibfield  {author} {\bibinfo {author} {\bibfnamefont {L.}~\bibnamefont
  {R\'ozsa}}, \bibinfo {author} {\bibfnamefont {E.}~\bibnamefont {Simon}},
  \bibinfo {author} {\bibfnamefont {K.}~\bibnamefont {Palot\'as}}, \bibinfo
  {author} {\bibfnamefont {L.}~\bibnamefont {Udvardi}},\ and\ \bibinfo {author}
  {\bibfnamefont {L.}~\bibnamefont {Szunyogh}},\ }\bibfield  {title} {\bibinfo
  {title} {{Complex magnetic phase diagram and skyrmion lifetime in an
  ultrathin film from atomistic simulations}},\ }\href
  {https://doi.org/10.1103/PhysRevB.93.024417} {\bibfield  {journal} {\bibinfo
  {journal} {Phys. Rev. B}\ }\textbf {\bibinfo {volume} {93}},\ \bibinfo
  {pages} {024417} (\bibinfo {year} {2016}{\natexlab{a}})}\BibitemShut
  {NoStop}%
\bibitem [{\citenamefont {Böttcher}\ \emph {et~al.}(2018)\citenamefont
  {Böttcher}, \citenamefont {Heinze}, \citenamefont {Egorov}, \citenamefont
  {Sinova},\ and\ \citenamefont {Dup{\'{e}}}}]{Bottcher2018}%
  \BibitemOpen
  \bibfield  {author} {\bibinfo {author} {\bibfnamefont {M.}~\bibnamefont
  {Böttcher}}, \bibinfo {author} {\bibfnamefont {S.}~\bibnamefont {Heinze}},
  \bibinfo {author} {\bibfnamefont {S.}~\bibnamefont {Egorov}}, \bibinfo
  {author} {\bibfnamefont {J.}~\bibnamefont {Sinova}},\ and\ \bibinfo {author}
  {\bibfnamefont {B.}~\bibnamefont {Dup{\'{e}}}},\ }\bibfield  {title}
  {\bibinfo {title} {{$B-T$ phase diagram of Pd/Fe/Ir(111) computed with
  parallel tempering Monte Carlo}},\ }\href
  {https://doi.org/10.1088/1367-2630/aae282} {\bibfield  {journal} {\bibinfo
  {journal} {New Journal of Physics}\ }\textbf {\bibinfo {volume} {20}},\
  \bibinfo {pages} {103014} (\bibinfo {year} {2018})}\BibitemShut {NoStop}%
\bibitem [{\citenamefont {R\'ozsa}\ \emph
  {et~al.}(2016{\natexlab{b}})\citenamefont {R\'ozsa}, \citenamefont {De\'ak},
  \citenamefont {Simon}, \citenamefont {Yanes}, \citenamefont {Udvardi},
  \citenamefont {Szunyogh},\ and\ \citenamefont {Nowak}}]{Rozsa2016PRL}%
  \BibitemOpen
  \bibfield  {author} {\bibinfo {author} {\bibfnamefont {L.}~\bibnamefont
  {R\'ozsa}}, \bibinfo {author} {\bibfnamefont {A.}~\bibnamefont {De\'ak}},
  \bibinfo {author} {\bibfnamefont {E.}~\bibnamefont {Simon}}, \bibinfo
  {author} {\bibfnamefont {R.}~\bibnamefont {Yanes}}, \bibinfo {author}
  {\bibfnamefont {L.}~\bibnamefont {Udvardi}}, \bibinfo {author} {\bibfnamefont
  {L.}~\bibnamefont {Szunyogh}},\ and\ \bibinfo {author} {\bibfnamefont
  {U.}~\bibnamefont {Nowak}},\ }\bibfield  {title} {\bibinfo {title}
  {{Skyrmions with Attractive Interactions in an Ultrathin Magnetic Film}},\
  }\href {https://doi.org/10.1103/PhysRevLett.117.157205} {\bibfield  {journal}
  {\bibinfo  {journal} {Phys. Rev. Lett.}\ }\textbf {\bibinfo {volume} {117}},\
  \bibinfo {pages} {157205} (\bibinfo {year} {2016}{\natexlab{b}})}\BibitemShut
  {NoStop}%
\bibitem [{\citenamefont {R\'ozsa}\ \emph {et~al.}(2017)\citenamefont
  {R\'ozsa}, \citenamefont {Palot\'as}, \citenamefont {De\'ak}, \citenamefont
  {Simon}, \citenamefont {Yanes}, \citenamefont {Udvardi}, \citenamefont
  {Szunyogh},\ and\ \citenamefont {Nowak}}]{Rozsa2017}%
  \BibitemOpen
  \bibfield  {author} {\bibinfo {author} {\bibfnamefont {L.}~\bibnamefont
  {R\'ozsa}}, \bibinfo {author} {\bibfnamefont {K.}~\bibnamefont {Palot\'as}},
  \bibinfo {author} {\bibfnamefont {A.}~\bibnamefont {De\'ak}}, \bibinfo
  {author} {\bibfnamefont {E.}~\bibnamefont {Simon}}, \bibinfo {author}
  {\bibfnamefont {R.}~\bibnamefont {Yanes}}, \bibinfo {author} {\bibfnamefont
  {L.}~\bibnamefont {Udvardi}}, \bibinfo {author} {\bibfnamefont
  {L.}~\bibnamefont {Szunyogh}},\ and\ \bibinfo {author} {\bibfnamefont
  {U.}~\bibnamefont {Nowak}},\ }\bibfield  {title} {\bibinfo {title}
  {{Formation and stability of metastable skyrmionic spin structures with
  various topologies in an ultrathin film}},\ }\href
  {https://doi.org/10.1103/PhysRevB.95.094423} {\bibfield  {journal} {\bibinfo
  {journal} {Phys. Rev. B}\ }\textbf {\bibinfo {volume} {95}},\ \bibinfo
  {pages} {094423} (\bibinfo {year} {2017})}\BibitemShut {NoStop}%
\bibitem [{\citenamefont {Schick}\ \emph {et~al.}(2021)\citenamefont {Schick},
  \citenamefont {Wei\ss{}enhofer}, \citenamefont {R\'ozsa},\ and\ \citenamefont
  {Nowak}}]{Schick2021}%
  \BibitemOpen
  \bibfield  {author} {\bibinfo {author} {\bibfnamefont {D.}~\bibnamefont
  {Schick}}, \bibinfo {author} {\bibfnamefont {M.}~\bibnamefont
  {Wei\ss{}enhofer}}, \bibinfo {author} {\bibfnamefont {L.}~\bibnamefont
  {R\'ozsa}},\ and\ \bibinfo {author} {\bibfnamefont {U.}~\bibnamefont
  {Nowak}},\ }\bibfield  {title} {\bibinfo {title} {{Skyrmions as
  quasiparticles: Free energy and entropy}},\ }\href
  {https://doi.org/10.1103/PhysRevB.103.214417} {\bibfield  {journal} {\bibinfo
   {journal} {Phys. Rev. B}\ }\textbf {\bibinfo {volume} {103}},\ \bibinfo
  {pages} {214417} (\bibinfo {year} {2021})}\BibitemShut {NoStop}%
\bibitem [{\citenamefont {Miranda}\ \emph {et~al.}(2022)\citenamefont
  {Miranda}, \citenamefont {Klautau}, \citenamefont {Bergman},\ and\
  \citenamefont {Petrilli}}]{Miranda2022}%
  \BibitemOpen
  \bibfield  {author} {\bibinfo {author} {\bibfnamefont {I.~P.}\ \bibnamefont
  {Miranda}}, \bibinfo {author} {\bibfnamefont {A.~B.}\ \bibnamefont
  {Klautau}}, \bibinfo {author} {\bibfnamefont {A.}~\bibnamefont {Bergman}},\
  and\ \bibinfo {author} {\bibfnamefont {H.~M.}\ \bibnamefont {Petrilli}},\
  }\bibfield  {title} {\bibinfo {title} {{Band filling effects on the emergence
  of magnetic skyrmions: Pd/Fe and Pd/Co bilayers on Ir(111)}},\ }\href
  {https://doi.org/10.1103/PhysRevB.105.224413} {\bibfield  {journal} {\bibinfo
   {journal} {Phys. Rev. B}\ }\textbf {\bibinfo {volume} {105}},\ \bibinfo
  {pages} {224413} (\bibinfo {year} {2022})}\BibitemShut {NoStop}%
\bibitem [{\citenamefont {Wei\ss{}enhofer}\ and\ \citenamefont
  {Nowak}(2019)}]{Weissenhofer2019}%
  \BibitemOpen
  \bibfield  {author} {\bibinfo {author} {\bibfnamefont {M.}~\bibnamefont
  {Wei\ss{}enhofer}}\ and\ \bibinfo {author} {\bibfnamefont {U.}~\bibnamefont
  {Nowak}},\ }\bibfield  {title} {\bibinfo {title} {{Orientation-dependent
  current-induced motion of skyrmions with various topologies}},\ }\href
  {https://doi.org/10.1103/PhysRevB.99.224430} {\bibfield  {journal} {\bibinfo
  {journal} {Phys. Rev. B}\ }\textbf {\bibinfo {volume} {99}},\ \bibinfo
  {pages} {224430} (\bibinfo {year} {2019})}\BibitemShut {NoStop}%
\bibitem [{\citenamefont {R{\'{o}}zsa}\ \emph {et~al.}(2020)\citenamefont
  {R{\'{o}}zsa}, \citenamefont {Wei{\ss}enhofer},\ and\ \citenamefont
  {Nowak}}]{Rozsa2020}%
  \BibitemOpen
  \bibfield  {author} {\bibinfo {author} {\bibfnamefont {L.}~\bibnamefont
  {R{\'{o}}zsa}}, \bibinfo {author} {\bibfnamefont {M.}~\bibnamefont
  {Wei{\ss}enhofer}},\ and\ \bibinfo {author} {\bibfnamefont {U.}~\bibnamefont
  {Nowak}},\ }\bibfield  {title} {\bibinfo {title} {{Spin waves in skyrmionic
  structures with various topological charges}},\ }\href
  {https://doi.org/10.1088/1361-648x/abc404} {\bibfield  {journal} {\bibinfo
  {journal} {Journal of Physics: Condensed Matter}\ }\textbf {\bibinfo {volume}
  {33}},\ \bibinfo {pages} {054001} (\bibinfo {year} {2020})}\BibitemShut
  {NoStop}%
\bibitem [{\citenamefont {Wei\ss{}enhofer}\ \emph {et~al.}(2021)\citenamefont
  {Wei\ss{}enhofer}, \citenamefont {R\'ozsa},\ and\ \citenamefont
  {Nowak}}]{Weissenhofer2021}%
  \BibitemOpen
  \bibfield  {author} {\bibinfo {author} {\bibfnamefont {M.}~\bibnamefont
  {Wei\ss{}enhofer}}, \bibinfo {author} {\bibfnamefont {L.}~\bibnamefont
  {R\'ozsa}},\ and\ \bibinfo {author} {\bibfnamefont {U.}~\bibnamefont
  {Nowak}},\ }\bibfield  {title} {\bibinfo {title} {{Skyrmion Dynamics at
  Finite Temperatures: Beyond Thiele's Equation}},\ }\href
  {https://doi.org/10.1103/PhysRevLett.127.047203} {\bibfield  {journal}
  {\bibinfo  {journal} {Phys. Rev. Lett.}\ }\textbf {\bibinfo {volume} {127}},\
  \bibinfo {pages} {047203} (\bibinfo {year} {2021})}\BibitemShut {NoStop}%
\bibitem [{\citenamefont {Bessarab}\ \emph {et~al.}(2015)\citenamefont
  {Bessarab}, \citenamefont {Uzdin},\ and\ \citenamefont
  {Jónsson}}]{BESSARAB2015335}%
  \BibitemOpen
  \bibfield  {author} {\bibinfo {author} {\bibfnamefont {P.~F.}\ \bibnamefont
  {Bessarab}}, \bibinfo {author} {\bibfnamefont {V.~M.}\ \bibnamefont
  {Uzdin}},\ and\ \bibinfo {author} {\bibfnamefont {H.}~\bibnamefont
  {Jónsson}},\ }\bibfield  {title} {\bibinfo {title} {{Method for finding
  mechanism and activation energy of magnetic transitions, applied to skyrmion
  and antivortex annihilation}},\ }\href
  {https://doi.org/https://doi.org/10.1016/j.cpc.2015.07.001} {\bibfield
  {journal} {\bibinfo  {journal} {Computer Physics Communications}\ }\textbf
  {\bibinfo {volume} {196}},\ \bibinfo {pages} {335} (\bibinfo {year}
  {2015})}\BibitemShut {NoStop}%
\bibitem [{\citenamefont {Cort\'{e}s-Ortu\~{n}o}\ \emph
  {et~al.}(2017)\citenamefont {Cort\'{e}s-Ortu\~{n}o}, \citenamefont {Wang},
  \citenamefont {Beg}, \citenamefont {Pepper}, \citenamefont {Bisotti},
  \citenamefont {Carey}, \citenamefont {Vousden}, \citenamefont {Kluyver},
  \citenamefont {Hovorka},\ and\ \citenamefont {Fangohr}}]{Cortes-Ortuno2017}%
  \BibitemOpen
  \bibfield  {author} {\bibinfo {author} {\bibfnamefont {D.}~\bibnamefont
  {Cort\'{e}s-Ortu\~{n}o}}, \bibinfo {author} {\bibfnamefont {W.}~\bibnamefont
  {Wang}}, \bibinfo {author} {\bibfnamefont {M.}~\bibnamefont {Beg}}, \bibinfo
  {author} {\bibfnamefont {R.~A.}\ \bibnamefont {Pepper}}, \bibinfo {author}
  {\bibfnamefont {M.-A.}\ \bibnamefont {Bisotti}}, \bibinfo {author}
  {\bibfnamefont {R.}~\bibnamefont {Carey}}, \bibinfo {author} {\bibfnamefont
  {M.}~\bibnamefont {Vousden}}, \bibinfo {author} {\bibfnamefont
  {T.}~\bibnamefont {Kluyver}}, \bibinfo {author} {\bibfnamefont
  {O.}~\bibnamefont {Hovorka}},\ and\ \bibinfo {author} {\bibfnamefont
  {H.}~\bibnamefont {Fangohr}},\ }\bibfield  {title} {\bibinfo {title} {Thermal
  stability and topological protection of skyrmions in nanotracks},\ }\href
  {https://doi.org/10.1038/s41598-017-03391-8} {\bibfield  {journal} {\bibinfo
  {journal} {Scientific Reports}\ }\textbf {\bibinfo {volume} {7}},\ \bibinfo
  {pages} {4060} (\bibinfo {year} {2017})}\BibitemShut {NoStop}%
\bibitem [{\citenamefont {Desplat}\ \emph {et~al.}(2018)\citenamefont
  {Desplat}, \citenamefont {Suess}, \citenamefont {Kim},\ and\ \citenamefont
  {Stamps}}]{Desplat2018}%
  \BibitemOpen
  \bibfield  {author} {\bibinfo {author} {\bibfnamefont {L.}~\bibnamefont
  {Desplat}}, \bibinfo {author} {\bibfnamefont {D.}~\bibnamefont {Suess}},
  \bibinfo {author} {\bibfnamefont {J.-V.}\ \bibnamefont {Kim}},\ and\ \bibinfo
  {author} {\bibfnamefont {R.~L.}\ \bibnamefont {Stamps}},\ }\bibfield  {title}
  {\bibinfo {title} {{Thermal stability of metastable magnetic skyrmions:
  Entropic narrowing and significance of internal eigenmodes}},\ }\href
  {https://doi.org/10.1103/PhysRevB.98.134407} {\bibfield  {journal} {\bibinfo
  {journal} {Phys. Rev. B}\ }\textbf {\bibinfo {volume} {98}},\ \bibinfo
  {pages} {134407} (\bibinfo {year} {2018})}\BibitemShut {NoStop}%
\bibitem [{\citenamefont {von Malottki}\ \emph {et~al.}(2019)\citenamefont {von
  Malottki}, \citenamefont {Bessarab}, \citenamefont {Haldar}, \citenamefont
  {Delin},\ and\ \citenamefont {Heinze}}]{vonMalottki2019}%
  \BibitemOpen
  \bibfield  {author} {\bibinfo {author} {\bibfnamefont {S.}~\bibnamefont {von
  Malottki}}, \bibinfo {author} {\bibfnamefont {P.~F.}\ \bibnamefont
  {Bessarab}}, \bibinfo {author} {\bibfnamefont {S.}~\bibnamefont {Haldar}},
  \bibinfo {author} {\bibfnamefont {A.}~\bibnamefont {Delin}},\ and\ \bibinfo
  {author} {\bibfnamefont {S.}~\bibnamefont {Heinze}},\ }\bibfield  {title}
  {\bibinfo {title} {{Skyrmion lifetime in ultrathin films}},\ }\href
  {https://doi.org/10.1103/PhysRevB.99.060409} {\bibfield  {journal} {\bibinfo
  {journal} {Phys. Rev. B}\ }\textbf {\bibinfo {volume} {99}},\ \bibinfo
  {pages} {060409} (\bibinfo {year} {2019})}\BibitemShut {NoStop}%
\bibitem [{\citenamefont {Muckel}\ \emph {et~al.}(2021)\citenamefont {Muckel},
  \citenamefont {von Malottki}, \citenamefont {Holl}, \citenamefont {Pestka},
  \citenamefont {Pratzer}, \citenamefont {Bessarab}, \citenamefont {Heinze},\
  and\ \citenamefont {Morgenstern}}]{Muckel2021}%
  \BibitemOpen
  \bibfield  {author} {\bibinfo {author} {\bibfnamefont {F.}~\bibnamefont
  {Muckel}}, \bibinfo {author} {\bibfnamefont {S.}~\bibnamefont {von
  Malottki}}, \bibinfo {author} {\bibfnamefont {C.}~\bibnamefont {Holl}},
  \bibinfo {author} {\bibfnamefont {B.}~\bibnamefont {Pestka}}, \bibinfo
  {author} {\bibfnamefont {M.}~\bibnamefont {Pratzer}}, \bibinfo {author}
  {\bibfnamefont {P.~F.}\ \bibnamefont {Bessarab}}, \bibinfo {author}
  {\bibfnamefont {S.}~\bibnamefont {Heinze}},\ and\ \bibinfo {author}
  {\bibfnamefont {M.}~\bibnamefont {Morgenstern}},\ }\bibfield  {title}
  {\bibinfo {title} {{Experimental identification of two distinct skyrmion
  collapse mechanisms}},\ }\href {https://doi.org/10.1038/s41567-020-01101-2}
  {\bibfield  {journal} {\bibinfo  {journal} {Nature Physics}\ }\textbf
  {\bibinfo {volume} {17}},\ \bibinfo {pages} {395} (\bibinfo {year}
  {2021})}\BibitemShut {NoStop}%
\bibitem [{\citenamefont {Hagemeister}\ \emph {et~al.}(2015)\citenamefont
  {Hagemeister}, \citenamefont {Romming}, \citenamefont {von Bergmann},
  \citenamefont {Vedmedenko},\ and\ \citenamefont
  {Wiesendanger}}]{Hagemeister2015}%
  \BibitemOpen
  \bibfield  {author} {\bibinfo {author} {\bibfnamefont {J.}~\bibnamefont
  {Hagemeister}}, \bibinfo {author} {\bibfnamefont {N.}~\bibnamefont
  {Romming}}, \bibinfo {author} {\bibfnamefont {K.}~\bibnamefont {von
  Bergmann}}, \bibinfo {author} {\bibfnamefont {E.~Y.}\ \bibnamefont
  {Vedmedenko}},\ and\ \bibinfo {author} {\bibfnamefont {R.}~\bibnamefont
  {Wiesendanger}},\ }\bibfield  {title} {\bibinfo {title} {{Stability of single
  skyrmionic bits}},\ }\href {https://doi.org/10.1038/ncomms9455} {\bibfield
  {journal} {\bibinfo  {journal} {Nature Communications}\ }\textbf {\bibinfo
  {volume} {6}},\ \bibinfo {pages} {8455} (\bibinfo {year} {2015})}\BibitemShut
  {NoStop}%
\bibitem [{\citenamefont {Desplat}\ \emph {et~al.}(2020)\citenamefont
  {Desplat}, \citenamefont {Vogler}, \citenamefont {Kim}, \citenamefont
  {Stamps},\ and\ \citenamefont {Suess}}]{Desplat2020}%
  \BibitemOpen
  \bibfield  {author} {\bibinfo {author} {\bibfnamefont {L.}~\bibnamefont
  {Desplat}}, \bibinfo {author} {\bibfnamefont {C.}~\bibnamefont {Vogler}},
  \bibinfo {author} {\bibfnamefont {J.-V.}\ \bibnamefont {Kim}}, \bibinfo
  {author} {\bibfnamefont {R.~L.}\ \bibnamefont {Stamps}},\ and\ \bibinfo
  {author} {\bibfnamefont {D.}~\bibnamefont {Suess}},\ }\bibfield  {title}
  {\bibinfo {title} {{Path sampling for lifetimes of metastable magnetic
  skyrmions and direct comparison with Kramers' method}},\ }\href
  {https://doi.org/10.1103/PhysRevB.101.060403} {\bibfield  {journal} {\bibinfo
   {journal} {Phys. Rev. B}\ }\textbf {\bibinfo {volume} {101}},\ \bibinfo
  {pages} {060403} (\bibinfo {year} {2020})}\BibitemShut {NoStop}%
\bibitem [{\citenamefont {Charalampidis}\ and\ \citenamefont
  {Barker}(2023)}]{charalampidis2023arxiv}%
  \BibitemOpen
  \bibfield  {author} {\bibinfo {author} {\bibfnamefont {I.}~\bibnamefont
  {Charalampidis}}\ and\ \bibinfo {author} {\bibfnamefont {J.}~\bibnamefont
  {Barker}},\ }\href@noop {} {\bibinfo {title} {{Metadynamics calculations of
  the effect of thermal spin fluctuations on skyrmion stability}}} (\bibinfo
  {year} {2023}),\ \Eprint {https://arxiv.org/abs/2310.03169} {arXiv:2310.03169
  [cond-mat.mtrl-sci]} \BibitemShut {NoStop}%
\bibitem [{\citenamefont {Bussi}\ and\ \citenamefont {Laio}(2020)}]{Bussi2020}%
  \BibitemOpen
  \bibfield  {author} {\bibinfo {author} {\bibfnamefont {G.}~\bibnamefont
  {Bussi}}\ and\ \bibinfo {author} {\bibfnamefont {A.}~\bibnamefont {Laio}},\
  }\bibfield  {title} {\bibinfo {title} {{Using metadynamics to explore complex
  free-energy landscapes}},\ }\href {https://doi.org/10.1038/s42254-020-0153-0}
  {\bibfield  {journal} {\bibinfo  {journal} {Nature Reviews Physics}\ }\textbf
  {\bibinfo {volume} {2}},\ \bibinfo {pages} {200} (\bibinfo {year}
  {2020})}\BibitemShut {NoStop}%
\bibitem [{\citenamefont {Lindner}\ \emph {et~al.}(2020)\citenamefont
  {Lindner}, \citenamefont {Bargsten}, \citenamefont {Kovarik}, \citenamefont
  {Friedlein}, \citenamefont {Harm}, \citenamefont {Krause},\ and\
  \citenamefont {Wiesendanger}}]{Lindner2020}%
  \BibitemOpen
  \bibfield  {author} {\bibinfo {author} {\bibfnamefont {P.}~\bibnamefont
  {Lindner}}, \bibinfo {author} {\bibfnamefont {L.}~\bibnamefont {Bargsten}},
  \bibinfo {author} {\bibfnamefont {S.}~\bibnamefont {Kovarik}}, \bibinfo
  {author} {\bibfnamefont {J.}~\bibnamefont {Friedlein}}, \bibinfo {author}
  {\bibfnamefont {J.}~\bibnamefont {Harm}}, \bibinfo {author} {\bibfnamefont
  {S.}~\bibnamefont {Krause}},\ and\ \bibinfo {author} {\bibfnamefont
  {R.}~\bibnamefont {Wiesendanger}},\ }\bibfield  {title} {\bibinfo {title}
  {{Temperature and magnetic field dependent behavior of atomic-scale skyrmions
  in Pd/Fe/Ir(111) nanoislands}},\ }\href
  {https://doi.org/10.1103/PhysRevB.101.214445} {\bibfield  {journal} {\bibinfo
   {journal} {Phys. Rev. B}\ }\textbf {\bibinfo {volume} {101}},\ \bibinfo
  {pages} {214445} (\bibinfo {year} {2020})}\BibitemShut {NoStop}%
\bibitem [{\citenamefont {Sch{\"a}ffer}\ \emph {et~al.}(2019)\citenamefont
  {Sch{\"a}ffer}, \citenamefont {R{\'o}zsa}, \citenamefont {Berakdar},
  \citenamefont {Vedmedenko},\ and\ \citenamefont
  {Wiesendanger}}]{Schaeffer2019}%
  \BibitemOpen
  \bibfield  {author} {\bibinfo {author} {\bibfnamefont {A.~F.}\ \bibnamefont
  {Sch{\"a}ffer}}, \bibinfo {author} {\bibfnamefont {L.}~\bibnamefont
  {R{\'o}zsa}}, \bibinfo {author} {\bibfnamefont {J.}~\bibnamefont {Berakdar}},
  \bibinfo {author} {\bibfnamefont {E.~Y.}\ \bibnamefont {Vedmedenko}},\ and\
  \bibinfo {author} {\bibfnamefont {R.}~\bibnamefont {Wiesendanger}},\
  }\bibfield  {title} {\bibinfo {title} {{Stochastic dynamics and pattern
  formation of geometrically confined skyrmions}},\ }\href
  {https://doi.org/10.1038/s42005-019-0176-y} {\bibfield  {journal} {\bibinfo
  {journal} {Communications Physics}\ }\textbf {\bibinfo {volume} {2}},\
  \bibinfo {pages} {72} (\bibinfo {year} {2019})}\BibitemShut {NoStop}%
\bibitem [{\citenamefont {Song}\ \emph {et~al.}(2021)\citenamefont {Song},
  \citenamefont {Kerber}, \citenamefont {Rothörl}, \citenamefont {Ge},
  \citenamefont {Raab}, \citenamefont {Seng}, \citenamefont {Brems},
  \citenamefont {Dittrich}, \citenamefont {Reeve}, \citenamefont {Wang},
  \citenamefont {Liu}, \citenamefont {Virnau},\ and\ \citenamefont
  {Kläui}}]{Song2021}%
  \BibitemOpen
  \bibfield  {author} {\bibinfo {author} {\bibfnamefont {C.}~\bibnamefont
  {Song}}, \bibinfo {author} {\bibfnamefont {N.}~\bibnamefont {Kerber}},
  \bibinfo {author} {\bibfnamefont {J.}~\bibnamefont {Rothörl}}, \bibinfo
  {author} {\bibfnamefont {Y.}~\bibnamefont {Ge}}, \bibinfo {author}
  {\bibfnamefont {K.}~\bibnamefont {Raab}}, \bibinfo {author} {\bibfnamefont
  {B.}~\bibnamefont {Seng}}, \bibinfo {author} {\bibfnamefont {M.~A.}\
  \bibnamefont {Brems}}, \bibinfo {author} {\bibfnamefont {F.}~\bibnamefont
  {Dittrich}}, \bibinfo {author} {\bibfnamefont {R.~M.}\ \bibnamefont {Reeve}},
  \bibinfo {author} {\bibfnamefont {J.}~\bibnamefont {Wang}}, \bibinfo {author}
  {\bibfnamefont {Q.}~\bibnamefont {Liu}}, \bibinfo {author} {\bibfnamefont
  {P.}~\bibnamefont {Virnau}},\ and\ \bibinfo {author} {\bibfnamefont
  {M.}~\bibnamefont {Kläui}},\ }\bibfield  {title} {\bibinfo {title}
  {{Commensurability between Element Symmetry and the Number of Skyrmions
  Governing Skyrmion Diffusion in Confined Geometries}},\ }\href
  {https://doi.org/https://doi.org/10.1002/adfm.202010739} {\bibfield
  {journal} {\bibinfo  {journal} {Advanced Functional Materials}\ }\textbf
  {\bibinfo {volume} {31}},\ \bibinfo {pages} {2010739} (\bibinfo {year}
  {2021})}\BibitemShut {NoStop}%
\bibitem [{\citenamefont {Ge}\ \emph {et~al.}(2023)\citenamefont {Ge},
  \citenamefont {Roth{\"o}rl}, \citenamefont {Brems}, \citenamefont {Kerber},
  \citenamefont {Gruber}, \citenamefont {Dohi}, \citenamefont {Kl{\"a}ui},\
  and\ \citenamefont {Virnau}}]{Ge2023}%
  \BibitemOpen
  \bibfield  {author} {\bibinfo {author} {\bibfnamefont {Y.}~\bibnamefont
  {Ge}}, \bibinfo {author} {\bibfnamefont {J.}~\bibnamefont {Roth{\"o}rl}},
  \bibinfo {author} {\bibfnamefont {M.~A.}\ \bibnamefont {Brems}}, \bibinfo
  {author} {\bibfnamefont {N.}~\bibnamefont {Kerber}}, \bibinfo {author}
  {\bibfnamefont {R.}~\bibnamefont {Gruber}}, \bibinfo {author} {\bibfnamefont
  {T.}~\bibnamefont {Dohi}}, \bibinfo {author} {\bibfnamefont {M.}~\bibnamefont
  {Kl{\"a}ui}},\ and\ \bibinfo {author} {\bibfnamefont {P.}~\bibnamefont
  {Virnau}},\ }\bibfield  {title} {\bibinfo {title} {{Constructing
  coarse-grained skyrmion potentials from experimental data with Iterative
  Boltzmann Inversion}},\ }\href {https://doi.org/10.1038/s42005-023-01145-9}
  {\bibfield  {journal} {\bibinfo  {journal} {Communications Physics}\ }\textbf
  {\bibinfo {volume} {6}},\ \bibinfo {pages} {30} (\bibinfo {year}
  {2023})}\BibitemShut {NoStop}%
\bibitem [{\citenamefont {Udvardi}\ \emph {et~al.}(2003)\citenamefont
  {Udvardi}, \citenamefont {Szunyogh}, \citenamefont {Palot\'as},\ and\
  \citenamefont {Weinberger}}]{Udvardi2003}%
  \BibitemOpen
  \bibfield  {author} {\bibinfo {author} {\bibfnamefont {L.}~\bibnamefont
  {Udvardi}}, \bibinfo {author} {\bibfnamefont {L.}~\bibnamefont {Szunyogh}},
  \bibinfo {author} {\bibfnamefont {K.}~\bibnamefont {Palot\'as}},\ and\
  \bibinfo {author} {\bibfnamefont {P.}~\bibnamefont {Weinberger}},\ }\bibfield
   {title} {\bibinfo {title} {{First-principles relativistic study of spin
  waves in thin magnetic films}},\ }\href
  {https://doi.org/10.1103/PhysRevB.68.104436} {\bibfield  {journal} {\bibinfo
  {journal} {Phys. Rev. B}\ }\textbf {\bibinfo {volume} {68}},\ \bibinfo
  {pages} {104436} (\bibinfo {year} {2003})}\BibitemShut {NoStop}%
\bibitem [{\citenamefont {Simon}\ \emph {et~al.}(2014)\citenamefont {Simon},
  \citenamefont {Palot\'as}, \citenamefont {R\'ozsa}, \citenamefont {Udvardi},\
  and\ \citenamefont {Szunyogh}}]{Simon2014}%
  \BibitemOpen
  \bibfield  {author} {\bibinfo {author} {\bibfnamefont {E.}~\bibnamefont
  {Simon}}, \bibinfo {author} {\bibfnamefont {K.}~\bibnamefont {Palot\'as}},
  \bibinfo {author} {\bibfnamefont {L.}~\bibnamefont {R\'ozsa}}, \bibinfo
  {author} {\bibfnamefont {L.}~\bibnamefont {Udvardi}},\ and\ \bibinfo {author}
  {\bibfnamefont {L.}~\bibnamefont {Szunyogh}},\ }\bibfield  {title} {\bibinfo
  {title} {{Formation of magnetic skyrmions with tunable properties in PdFe
  bilayer deposited on Ir(111)}},\ }\href
  {https://doi.org/10.1103/PhysRevB.90.094410} {\bibfield  {journal} {\bibinfo
  {journal} {Phys. Rev. B}\ }\textbf {\bibinfo {volume} {90}},\ \bibinfo
  {pages} {094410} (\bibinfo {year} {2014})}\BibitemShut {NoStop}%
\bibitem [{\citenamefont {{Laio}}\ and\ \citenamefont
  {{Parrinello}}(2002)}]{Laio2002}%
  \BibitemOpen
  \bibfield  {author} {\bibinfo {author} {\bibfnamefont {A.}~\bibnamefont
  {{Laio}}}\ and\ \bibinfo {author} {\bibfnamefont {M.}~\bibnamefont
  {{Parrinello}}},\ }\bibfield  {title} {\bibinfo {title} {{{Escaping
  free-energy minima}}},\ }\href {https://doi.org/10.1073/pnas.202427399}
  {\bibfield  {journal} {\bibinfo  {journal} {Proc. Nat. Acad. Sci.}\ }\textbf
  {\bibinfo {volume} {99}},\ \bibinfo {pages} {12562} (\bibinfo {year}
  {2002})}\BibitemShut {NoStop}%
\bibitem [{\citenamefont {{Barducci}}\ \emph {et~al.}(2008)\citenamefont
  {{Barducci}}, \citenamefont {{Bussi}},\ and\ \citenamefont
  {{Parrinello}}}]{Barducci2008}%
  \BibitemOpen
  \bibfield  {author} {\bibinfo {author} {\bibfnamefont {A.}~\bibnamefont
  {{Barducci}}}, \bibinfo {author} {\bibfnamefont {G.}~\bibnamefont
  {{Bussi}}},\ and\ \bibinfo {author} {\bibfnamefont {M.}~\bibnamefont
  {{Parrinello}}},\ }\bibfield  {title} {\bibinfo {title} {{{Well-Tempered
  Metadynamics: A Smoothly Converging and Tunable Free-Energy Method}}},\
  }\href {https://doi.org/10.1103/PhysRevLett.100.020603} {\bibfield  {journal}
  {\bibinfo  {journal} {Phys. Rev. Lett.}\ }\textbf {\bibinfo {volume} {100}},\
  \bibinfo {eid} {020603} (\bibinfo {year} {2008})}\BibitemShut {NoStop}%
\bibitem [{\citenamefont {Nagyfalusi}\ \emph {et~al.}(2019)\citenamefont
  {Nagyfalusi}, \citenamefont {Udvardi},\ and\ \citenamefont
  {Szunyogh}}]{Nagyfalusi2019}%
  \BibitemOpen
  \bibfield  {author} {\bibinfo {author} {\bibfnamefont {B.}~\bibnamefont
  {Nagyfalusi}}, \bibinfo {author} {\bibfnamefont {L.}~\bibnamefont
  {Udvardi}},\ and\ \bibinfo {author} {\bibfnamefont {L.}~\bibnamefont
  {Szunyogh}},\ }\bibfield  {title} {\bibinfo {title} {{Metadynamics study of
  the temperature dependence of magnetic anisotropy and spin-reorientation
  transitions in ultrathin films}},\ }\href
  {https://doi.org/10.1103/PhysRevB.100.174429} {\bibfield  {journal} {\bibinfo
   {journal} {Phys. Rev. B}\ }\textbf {\bibinfo {volume} {100}},\ \bibinfo
  {pages} {174429} (\bibinfo {year} {2019})}\BibitemShut {NoStop}%
\bibitem [{\citenamefont {Nagyfalusi}\ \emph {et~al.}(2020)\citenamefont
  {Nagyfalusi}, \citenamefont {Udvardi}, \citenamefont {Szunyogh},\ and\
  \citenamefont {R\'ozsa}}]{Nagyfalusi2020}%
  \BibitemOpen
  \bibfield  {author} {\bibinfo {author} {\bibfnamefont {B.}~\bibnamefont
  {Nagyfalusi}}, \bibinfo {author} {\bibfnamefont {L.}~\bibnamefont {Udvardi}},
  \bibinfo {author} {\bibfnamefont {L.}~\bibnamefont {Szunyogh}},\ and\
  \bibinfo {author} {\bibfnamefont {L.}~\bibnamefont {R\'ozsa}},\ }\bibfield
  {title} {\bibinfo {title} {{Spin reorientation transition in an ultrathin Fe
  film on W(110) induced by Dzyaloshinsky-Moriya interactions}},\ }\href
  {https://doi.org/10.1103/PhysRevB.102.134413} {\bibfield  {journal} {\bibinfo
   {journal} {Phys. Rev. B}\ }\textbf {\bibinfo {volume} {102}},\ \bibinfo
  {pages} {134413} (\bibinfo {year} {2020})}\BibitemShut {NoStop}%
\bibitem [{\citenamefont {Polyakov}\ and\ \citenamefont
  {Belavin}(1975)}]{Polyakov1975}%
  \BibitemOpen
  \bibfield  {author} {\bibinfo {author} {\bibfnamefont {A.~M.}\ \bibnamefont
  {Polyakov}}\ and\ \bibinfo {author} {\bibfnamefont {A.~A.}\ \bibnamefont
  {Belavin}},\ }\bibfield  {title} {\bibinfo {title} {{{Metastable States of
  Two-Dimensional Isotropic Ferromagnets}}},\ }\href@noop {} {\bibfield
  {journal} {\bibinfo  {journal} {JETP Lett.}\ }\textbf {\bibinfo {volume}
  {22}},\ \bibinfo {pages} {245} (\bibinfo {year} {1975})}\BibitemShut
  {NoStop}%
\bibitem [{\citenamefont {Berg}\ and\ \citenamefont
  {L\"{u}scher}(1981)}]{Berg1981}%
  \BibitemOpen
  \bibfield  {author} {\bibinfo {author} {\bibfnamefont {B.}~\bibnamefont
  {Berg}}\ and\ \bibinfo {author} {\bibfnamefont {M.}~\bibnamefont
  {L\"{u}scher}},\ }\bibfield  {title} {\bibinfo {title} {{Definition and
  statistical distributions of a topological number in the lattice O(3)
  $\sigma$-model}},\ }\href {https://doi.org/10.1016/0550-3213(81)90568-x}
  {\bibfield  {journal} {\bibinfo  {journal} {Nuclear Physics B}\ }\textbf
  {\bibinfo {volume} {190}},\ \bibinfo {pages} {412} (\bibinfo {year}
  {1981})}\BibitemShut {NoStop}%
\bibitem [{\citenamefont {Dama}\ \emph {et~al.}(2014)\citenamefont {Dama},
  \citenamefont {Parrinello},\ and\ \citenamefont {Voth}}]{Dama2014}%
  \BibitemOpen
  \bibfield  {author} {\bibinfo {author} {\bibfnamefont {J.~F.}\ \bibnamefont
  {Dama}}, \bibinfo {author} {\bibfnamefont {M.}~\bibnamefont {Parrinello}},\
  and\ \bibinfo {author} {\bibfnamefont {G.~A.}\ \bibnamefont {Voth}},\
  }\bibfield  {title} {\bibinfo {title} {{Well-Tempered Metadynamics Converges
  Asymptotically}},\ }\href {https://doi.org/10.1103/PhysRevLett.112.240602}
  {\bibfield  {journal} {\bibinfo  {journal} {Phys. Rev. Lett.}\ }\textbf
  {\bibinfo {volume} {112}},\ \bibinfo {pages} {240602} (\bibinfo {year}
  {2014})}\BibitemShut {NoStop}%
\bibitem [{\citenamefont {Nagyfalusi}\ \emph {et~al.}(2022)\citenamefont
  {Nagyfalusi}, \citenamefont {Udvardi},\ and\ \citenamefont
  {Szunyogh}}]{Nagyfalusi_2022}%
  \BibitemOpen
  \bibfield  {author} {\bibinfo {author} {\bibfnamefont {B.}~\bibnamefont
  {Nagyfalusi}}, \bibinfo {author} {\bibfnamefont {L.}~\bibnamefont
  {Udvardi}},\ and\ \bibinfo {author} {\bibfnamefont {L.}~\bibnamefont
  {Szunyogh}},\ }\bibfield  {title} {\bibinfo {title} {{Magnetic ground state
  of supported monatomic Fe chains from first principles}},\ }\href
  {https://doi.org/10.1088/1361-648X/ac8260} {\bibfield  {journal} {\bibinfo
  {journal} {Journal of Physics: Condensed Matter}\ }\textbf {\bibinfo {volume}
  {34}},\ \bibinfo {pages} {395803} (\bibinfo {year} {2022})}\BibitemShut
  {NoStop}%
\end{thebibliography}%

\end{document}